\documentclass[twocolumn,epjc3]{svjour3}  
\smartqed   
\RequirePackage{xcolor}
\RequirePackage[colorlinks,linkcolor=blue,citecolor=blue,urlcolor=blue]{hyperref} 
\RequirePackage{newtxtext,newtxmath}
\RequirePackage{doi}
\RequirePackage{graphicx}
\RequirePackage{multirow}
\RequirePackage{array}
\RequirePackage{mathtools}
\RequirePackage{amsmath}
\RequirePackage{amsfonts}
\RequirePackage{tikz}
\usetikzlibrary{positioning}

\RequirePackage{bm}
\allowdisplaybreaks

\RequirePackage[numbers,sort&compress]{natbib}

\newcommand\tk{\tikz\fill[scale=0.25](0,.35) -- (.25,0) -- (1,.7) -- (.25,.15) -- cycle;}

\newcolumntype{M}[1]{>{\hbox to #1\bgroup\hss$}l<{$\egroup}}

\makeatletter
\newcommand\@brcolwidth{0.67em}
\newenvironment{brmatrix}{%
    \left[%
    \hskip-\arraycolsep
    \new@ifnextchar[\@brarray{\@brarray[\@brcolwidth]}%
}{%
    \endarray
    \hskip -\arraycolsep
    \right]%
}
\def\@brarray[#1]{\array{r*\c@MaxMatrixCols {M{#1}}}}
\makeatother

\newcommand{\im}{\operatorname{Im}}
\newcommand{\disc}{\operatorname{Disc}}
\newcommand{\dsz}{D_{s0}^*(2317)}
\newcommand{\dso}{D_{s1}(2460)}
\newcommand{\dzero}{D_0^*(2300)}
\newcommand{\dz}{D_0^*}
\newcommand{\done}{D_1(2430)}
\newcommand{\mev}{{\rm MeV}}
\newcommand{\gev}{{\rm GeV}}

\newcommand{\gdp}{g_{D\pi}^r}
\newcommand{\gdeta}{g_{D\eta}^r}
\newcommand{\gdsk}{g_{D_s\bar K}^r}
\journalname{Eur. Phys. J. C}
\begin{document}

\title{Can the two-pole structure of the {\boldmath$D_0^*(2300)$} be understood
from recent lattice data?
}

\author{Anuvind~Asokan\thanksref{e1,addr1}\and
        Meng-Na~Tang\thanksref{e6,addr2,addr3}
        \and
        Feng-Kun~Guo\thanksref{e2,addr2,addr3,addr4}
         \and
        Christoph~Hanhart\thanksref{e3,addr1} \and
        Yuki~Kamiya\thanksref{e4,addr5} \and
        Ulf-G.~Mei\ss ner\thanksref{e5,addr5,addr1,addr6} 
}

\thankstext{e1}{e-mail: a.asokan@fz-juelich.de}
\thankstext{e2}{e-mail: fkguo@itp.ac.cn}
\thankstext{e3}{e-mail: c.hanhart@fz-juelich.de}
\thankstext{e4}{e-mail: kamiya@hiskp.uni-bonn.de}
\thankstext{e5}{e-mail: meissner@hiskp.uni-bonn.de}
\thankstext{e6}{e-mail: tangmengna@mail.itp.ac.cn}

\institute{Institute for Advanced Simulation and Institut für Kernphysik, Forschungszentrum Jülich, D-52425 J\"ulich, Germany \label{addr1}
\and
           CAS Key Laboratory of Theoretical Physics, Institute of Theoretical Physics,
Chinese Academy of Sciences, Beijing 100190, China
\label{addr2}
           \and
           School of Physical Sciences, University of Chinese Academy of Sciences, Beijing 100049, China \label{addr3}
           \and
           Peng Huanwu Collaborative Center for Research and Education, Beihang University, Beijing 100191, China \label{addr4}
           \and
           Helmholtz-Institut f\"ur Strahlen- und Kernphysik and Bethe Center for Theoretical Physics, Universit\"at Bonn, D-53115 Bonn, Germany \label{addr5}
           \and
           Tbilisi State University, 0186 Tbilisi, Georgia \label{addr6}
}

\date{Received: date / Accepted: date}

\maketitle

\begin{abstract}
It was demonstrated in a series of papers employing unitarized chiral perturbation theory that the phenomenology of the scalar open-charm state, the 
$D_0^*(2300)$, can be understood as
the interplay of two poles, corresponding to two scalar-isospin doublet states with different SU(3) flavor content. Within this formalism the lightest open charm
positive parity states emerge as being dynamically generated from 
the scattering of the Goldstone-boson octet off $D$ mesons, a picture that
at the same time solves various problems that the experimental observations
posed. However, in recent lattice studies of $D\pi$ scattering at different
pion masses only one pole was reported in the $D_0^*$ channel, while it was
not possible to extract reliable parameters of a second pole from the lattice data. In this paper
we demonstrate how this seeming contradiction can be understood and that
imposing SU(3) constraints on the fitting amplitudes allows one to
extract information on the second pole from the lattice data with minimal bias.
The results may also be regarded as a showcase how 
approximate symmetries can be imposed in the $K$-matrix formalism to reduce the number of parameters.

\end{abstract}

\section{Introduction}
\label{intro}

The discovery of the charm-strange mesons $\dsz$~\cite{BaBar:2003oey} and $\dso$~\cite{CLEO:2003ggt} with masses significantly lower than the predictions for the lowest-lying scalar and axial-vector $c\bar s$ mesons from the quark model (see, e.g., Ref.~\cite{Godfrey:1985xj}) in 2003 lead to intensive discussions on their nature. 
Closely related to these two hadrons, there were observations of broad bumps in the $D\pi$ and $D^*\pi$ invariant mass distributions in $B$ decays by BaBar, Belle and LHCb Collaborations~\cite{Belle:2003nsh,BaBar:2009pnd,LHCb:2015klp,LHCb:2015tsv}. The bumps were fitted using a
 Breit-Wigner (BW) parametrization with energy-dependent widths, assuming the existence of one broad scalar (axial-vector) resonance coupled to $D^{(*)}\pi$; accordingly, such fits led to the $\dzero$ and $\done$ entries that are listed in the Review of Particle Physics (RPP)~\cite{Workman:2022ynf}:
\begin{align}
    M_{\dz} &= 2343\pm 10, & \Gamma_{\dz} &= 229\pm 16 , \notag\\
    M_{D_1} &= 2422.1\pm 0.6, & \Gamma_{D_1} &= 411.8\pm 0.6 \ ,
\end{align}
where all numbers are given in units of MeV.

However, the use of a BW form is not justified in these cases as constraints from chiral symmetry
and coupled channel effects are not taken into account, see e.g.~\cite{Du:2019oki}. 
Those are automatically built into unitarized chiral perturbation theory (UChPT),
where all calculations find two $\dz$ mesons and two $D_1$ mesons in the same energy region as the $\dzero$ and $\done$, respectively~\cite{Kolomeitsev:2003ac,Guo:2006fu,Guo:2006rp,Guo:2009ct,Albaladejo:2016lbb,Du:2017zvv,
Guo:2019o4,Lutz:2022enz}.\footnote{What is meant here is that there are two states coupling to the $\pi D^{(*)}$ channel
predominantly in $S$-wave, instead of only one $D_0^*(2300)/D_1(2430)$. Clearly, in addition to these states there
is also the narrow $D_1(2420)$, decaying into $\pi D^*$ predominantly in $D$-wave (up to heavy quark spin symmetry
violating contributions) which has a width of about 30 MeV.}
All these works tell a qualitatively coherent story, although e.g. the role of
left-hands cuts still needs be agreed upon~\cite{Lutz:2022enz,Korpa:2022voo}.
For instance, the parameters in the UChPT amplitude used in Refs.~\cite{Albaladejo:2016lbb,Du:2017zvv} are fixed from fitting to the results of a set of $S$-wave charmed-meson--light-pseudoscalar-meson ($D\Phi$) scattering lengths computed using lattice quantum chromodynamics (QCD)~\cite{Liu:2012zya}. 
The two $\dz$ poles in the UChPT amplitude of Refs.~\cite{Liu:2012zya,Albaladejo:2016lbb,Du:2017zvv} are located at $2105_{-8}^{+6}- i 102_{-11}^{+10}$~\mev\ and $2451_{-26}^{+35}-i 134_{-8}^{+7}$~\mev.
And it was demonstrated in Refs.~\cite{Du:2017zvv,Du:2019oki} that the amplitudes are consistent with the LHCb data of the angular moment distributions from three-body $B$ meson decays: $B^- \rightarrow D^{+} \pi^- \pi^-$~\cite{LHCb:2016lxy}, $B_s^0 \rightarrow \bar{D}^0 K^- \pi^{+}$~\cite{LHCb:2014ioa}, $B^0 \rightarrow \bar{D}^0 \pi^- \pi^{+}$~\cite{LHCb:2015klp}, $B^- \rightarrow D^{+} \pi^- K^-$~\cite{LHCb:2015eqv}, and $B^0 \rightarrow \bar{D}^0 \pi^- K^{+}$~\cite{LHCb:2015tsv}. For a review on two-pole
structures in QCD, see~\cite{Meissner:2020khl}.

In seeming disagreement to these findings, the lattice QCD analysis of the $D\pi$-$D\eta$-$D_s\bar K$ 
coupled channel  system by the Hadron Spectrum Collaboration (HadSpec) in Ref.~\cite{Moir:2016srx} reported 
only one $\dz$ state just below the $D\pi$ threshold, with the pion mass of about 391~\mev.
In this paper, we will discuss whether the higher $\dz$ pole is consistent with the lattice 
data, and propose a $K$-matrix formalism constrained with the SU(3) flavor symmetry that can 
be used in analyzing coupled-channel lattice data. 

\section{Analysis of the Amplitude from the Lattice study}
\label{sec:lanalysis}
In Ref.~\cite{Moir:2016srx} lattice data for the strangeness zero, isospin-1/2 channel at a pion mass of 
about 391~\mev\ 
were presented and analyzed with a sizable set of $K$-matrix parametrizations of the kind
\begin{align}
    K_{ij} = \frac{\left(g^{(0)}_i {+}\,g^{(1)}_i s\right)\left(g^{(0)}_j {+}\,g^{(1)}_j s\right)}{m^2 - s} 
    +\,\gamma^{(0)}_{ij} +\,\gamma^{(1)}_{ij} s,
    \label{eq:Kdef}
\end{align}
where $i$ and $j$ label the different reaction channels and $m$, $g^{(n)}_i$\ and $\gamma^{(n)}_{ij}$\ are real parameters to be determined 
in the fit to the lattice data. 
From this, the $T$-matrix for the $S$-wave coupled-channel ($D\pi$-$D\eta$-$D_s\bar K$) scattering is given by
\begin{equation}
T(s) = -16\pi\, T_K(s),
\end{equation}
with $T_K(s)$ defined as
\begin{align}
    T_K^{-1}(s)_{ij} = K^{-1}(s)_{ij} + \left(I^{(i)}_{\rm CM}(s)-I^{(i)}_{\rm CM}(m^2)\right)\delta_{ij},
    \label{eq:Tmat}
\end{align}
where the second term on the right-hand side contains the Chew-Mandelstam function, subtracted
at the $K$-matrix pole parameter $m$. It is given by
\begin{eqnarray}
     I^{(i)}_{\rm CM}(s) &=& \frac{\rho_i(s)}{\pi} \log\left[\frac{\xi_i(s)+\rho_i(s)}{\xi_i(s)-\rho_i(s)}\right] \nonumber \\ 
  & &  - \frac{\xi_i(s)}{\pi} 
     \frac{m^{(i)}_2-m^{(i)}_1}{m^{(i)}_1 + m^{(i)}_2} \log\frac{m^{(i)}_2}{m^{(i)}_1},
     \label{eq:chew}
 \end{eqnarray}
with
\begin{align}
\xi_i(s) &= 1 - \frac{\left(m^{(i)}_1+m^{(i)}_2\right)^2}{s}, \\
\rho_i^2(s) & =  \xi_i(s)
    \left( 1 -\frac{( m^{(i)}_1 - m^{(i)}_2)^2}{s} \right),\label{eq:rho}
\end{align}
where $m^{(i)}_1$\ and $m^{(i)}_2$\ are the masses of the two particles in channel $i$\ 
and $s$\ is the centre-of-mass (c.m.) energy squared.
The imaginary part of $T_K^{-1}(s)_{ij}$ is then given by the phase-space factor
$-\delta_{ij}\rho_{j}\theta\left(\sqrt{s}-m^{(i)}_1 - m^{(i)}_2\right)$, which automatically ensures the unitarity of the $S$-matrix.

The nine parametrizations presented in Ref.~\cite{Moir:2016srx} differed
by the set of parameters that was allowed to vary in the course of the fit.
The 
parameters present in the different amplitudes along with their reduced $\chi^2$
values from energy level fits performed in Ref.~\cite{Moir:2016srx} are given in 
Table~\ref{tab:pvar}. 

\begin{table*}
    \centering
    \caption{The parametrizations used in the analysis in Ref.~\cite{Moir:2016srx}. The check mark denotes a free parameter 
    and ``-'' implies the parameter is fixed to zero. The channels are denoted with increasing 
    threshold energies, with $1=D\pi$, $2=D\eta$ and $3=D_s\Bar{K}$.}
    \begin{tabular*}{\textwidth}{@{\extracolsep{\fill}}*{21}{c}@{}}
    \hline\noalign{\smallskip}
        \multirow{2}{7em}{Parametrization} & \multirow{2}{1em}{$m$} & \multicolumn{3}{c}{$g_i^{(0)}$} & \multicolumn{3}{c}{$g_i^{(1)}$} & 
        \multicolumn{6}{c}{$\gamma_{ij}^{(0)}$}  &
        \multicolumn{6}{c}{$\gamma_{ij}^{(1)}$} & \multirow{2}{3.5em}{$\chi^2/\mbox{dof}$}\\
        
         & & 1 & 2 & 3 & 1 & 2 & 3 & 11 & 12 & 13 & 22 & 23 & 33 & 11 & 12 & 13 & 22 & 23 & 33 & \\
         \noalign{\smallskip}\hline\noalign{\smallskip}
         Amplitude 1 & \tk & \tk & \tk & \tk & - & - & - &  \tk &  \tk &  \tk &  \tk &  - &  \tk &  - & - &  - &  - &  - &  - & 1.76\\
         Amplitude 2 & \tk & \tk & \tk & \tk & - & - & - & \tk & \tk &  - &  \tk &  - &  \tk &  - & - &  - &  - &  - &  - & 1.71\\
         Amplitude 3 & \tk & \tk & \tk & \tk & - & - & - &  \tk &  - &  - &  \tk &  - &  \tk &  - &  - &  - &  - &  - &  - & 1.76\\
         Amplitude 4 & \tk & \tk & \tk & \tk & - & - & - &  - &  - &  - &  \tk &  - &  - &  \tk & - &  - &  - &  - &  - & 1.78\\
         Amplitude 5 & \tk & \tk & \tk & \tk & - & - & - &  - &  - &  - &  - &  - &  - &  \tk & - &  - &  \tk &  - &  \tk  & 1.89\\
         Amplitude 6 & \tk & \tk & \tk & \tk & \tk & - & - & \tk &  - &  - & \tk &  - &  \tk &  - & - &  - &  - &  - &  -  & 1.63\\
         Amplitude 7 & \tk & \tk & \tk & \tk & \tk & - & - & \tk &  - &  - & \tk &  - &  - &  - & - &  - &  - &  - &  -  & 1.68\\
         Amplitude 8 & \tk & \tk & \tk & \tk & \tk & - & \tk & \tk &  - &  - & \tk &  - & \tk &  - & - &  - &  - &  - &  -  & 1.68\\
         Amplitude 9 & \tk & \tk & \tk & \tk & \tk & \tk & - &  \tk &  - &  - & \tk &  - & \tk &  - & - &  - &  - &  - &  -  & 1.66\\
         \noalign{\smallskip}\hline
    \end{tabular*}
    \label{tab:pvar}
\end{table*}

\subsection{Pole Search}
\label{subsec:polesearch}
\begin{figure}
    \centering
    \includegraphics[scale=0.9]{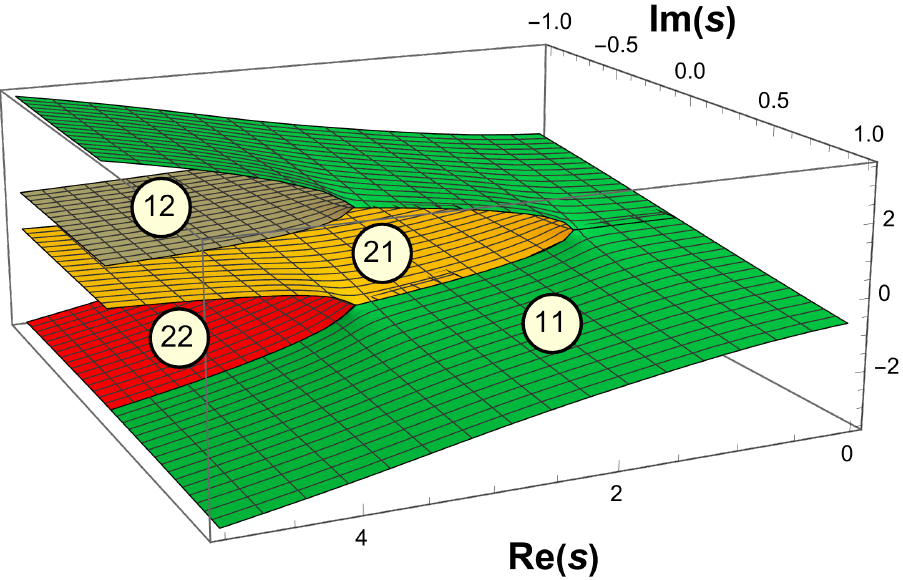}
    \caption{Illustration for the sheet labeling in the case of
    two channels.}
    \label{fig:twochannelsheets}
\end{figure}

The $T$-matrix is analytic over the whole complex energy plane
except for poles and branch cuts along the real axis due to kinematic (right-hand
cuts) and dynamic singularities (left-hand cuts). Dynamic singularities (left-hand cuts) are associated with the interactions in the crossed channels.
Since those are usually distant, one assumes that their effect can be captured by polynomial
terms allowed in the parametrization of the $K$-matrix used.
Right-hand cuts start from branch points that appear whenever a channel opens.
Accordingly, at each threshold the number of Riemann sheets of the complex energy (or $s$) plane gets doubled.
Thus, the three-channel case studied here leads to eight Riemann sheets. The sheets are labeled as 
shown in Table~\ref{tab:sheet}, where the thresholds are arranged with increasing energies $1=D\pi$, 
$2=D\eta$ and $3=D_s\Bar{K}$. For illustration we show in Fig.~\ref{fig:twochannelsheets} the 
analogous labeling for two channels. See Fig.~3 of Ref.~\cite{Mai:2022eur}
for the three-channel case.

\begin{table*}
    \centering
    \caption{The notation of the Riemann sheets with the sign of the imaginary part of the c.m. momentum of each channel.}
    \begin{tabular*}{\textwidth}{@{\extracolsep{\fill}}cccc@{}}
    \hline\noalign{\smallskip}
         Riemann sheet& \multicolumn{3}{c}{Sign of imaginary part of channel momentum} \\
    \noalign{\smallskip}\hline\noalign{\smallskip}
         RS111&$\im(p_1)~>~0$ & $\im(p_2)~>~0$ & $\im(p_3)~>~0$  \\
         RS211& $\im(p_1)~<~0$ & $\im(p_2)~>~0$ & $\im(p_3)~>~0$  \\
         RS221& $\im(p_1)~<~0$ & $\im(p_2)~<~0$ & $\im(p_3)~>~0$  \\
         RS222& $\im(p_1)~<~0$ & $\im(p_2)~<~0$ & $\im(p_3)~<~0$  \\
         RS121& $\im(p_1)~>~0$ & $\im(p_2)~<~0$ & $\im(p_3)~>~0$  \\
         RS112& $\im(p_1)~>~0$ & $\im(p_2)~>~0$ & $\im(p_3)~<~0$  \\
         RS212& $\im(p_1)~<~0$ & $\im(p_2)~>~0$ & $\im(p_3)~<~0$  \\
         RS122& $\im(p_1)~>~0$ & $\im(p_2)~<~0$ & $\im(p_3)~<~0$  \\
         \noalign{\smallskip}\hline
    \end{tabular*}
    \label{tab:sheet}
\end{table*}

The poles correspond to bound states or resonances 
depending on their location on the Riemann sheets. Bound states correspond to poles 
on the physical sheet below the lowest threshold energy and resonances are poles in the 
complex plane of the
unphysical sheets (in addition there are virtual state poles, located on the
real axis of unphysical sheets, but those do not play a role in this
work). The poles on the sheets closest to the physical sheet have the 
strongest influence on the scattering amplitude. In the current notation 
sheets RS211, RS221, and RS222 
would be directly connected to the physical sheet, i.e., RS111, above the respective thresholds
(c.f. Fig.~\ref{fig:twochannelsheets}). 
The poles of the $T$-matrix 
are given by the zeroes of the determinant of the matrix in 
Eq.~\eqref{eq:Tmat}, i.e.,
\begin{align}
    {\rm det}\left(K^{-1}(s) + (I_{\rm CM}(s)-I_{\rm CM}(m^2))\right) = 0.
    \label{eq:det}
\end{align}

The unphysical sheets can be accessed by adding the discontinuity 
across the branch cut to Eq.~\eqref{eq:Tmat}. Via the
Schwarz reflection principle the discontinuity across 
the branch cut is related to the imaginary part of the amplitude by
\begin{align}
    \disc\left[T_K(s)\right] = T_K(s {+} i\epsilon) - T_K(s - i\epsilon)= 2i\,\im[T_K(s {+} i\epsilon)],
    \label{discim}
\end{align}
where $\im[T_K(s {+} i\epsilon)]$ needs to be understood as the analytic continuation of the imaginary part of the amplitude on the real axis above threshold.

Crossing from the physical sheet (RS111) to any 
sheet can be done by
\begin{align}
    T_{K,X}^{-1}(s) = T_K^{-1}(s)  +  \disc_{X}[T_K^{-1}(s)], 
\end{align}
where the subscript $X$\ stands for the sheet number and $\disc$ is a $3\times3$
matrix containing the relevant discontinuities needed for the sheet transition, 
e.g. for the transition from RS111 to RS211 we employ
\begin{equation}
    \disc_{211}T_K^{-1} = 2i\begin{brmatrix}[2em]
            -\rho_{1} & 0 & 0\\
             0 & 0 & 0\\
              0 & 0 & 0
           \end{brmatrix} ,
    \end{equation}
    and for RS111 to RS221 
    \begin{equation}
    \disc_{221}T_K^{-1} = 2i\begin{brmatrix}[2em]
            -\rho_{1} & 0 & 0\\
             0 & -\rho_{2} & 0 \\
              0 & 0 & 0
           \end{brmatrix}    . 
\end{equation}
This prescription is straightforwardly generalized to arbitrary transitions 
between sheets.

At a pion mass of about 391~\mev, the lowest pole in the studied channel turns out to
be a bound state, accordingly located on sheet RS111~\cite{Moir:2016srx}; the same conclusion was reached in UChPT in Ref.~\cite{Albaladejo:2016lbb}.
This pole was found in the fits of all 9 parametrizations employed by the Hadron Spectrum Collaboration~\cite{Moir:2016srx}. 
At the same time, additional poles were 
found on sheets RS211, RS221, and RS222. These additional poles were found for 
almost all amplitude paramterizations employed in Ref.~\cite{Moir:2016srx}, which were,
however, not reported in the publication since they not only scatter very
much, but also are in parts located outside the energy region where the 
fit was performed.
Table~\ref{tab:poles} shows the pole values found from the search with 
the corresponding sheets from the different amplitude parametrizations.
The $1\sigma$\ uncertainties of the pole values were 
calculated by the bootstrap method.  

\begin{figure}
    \centering
    \includegraphics[scale=0.44]{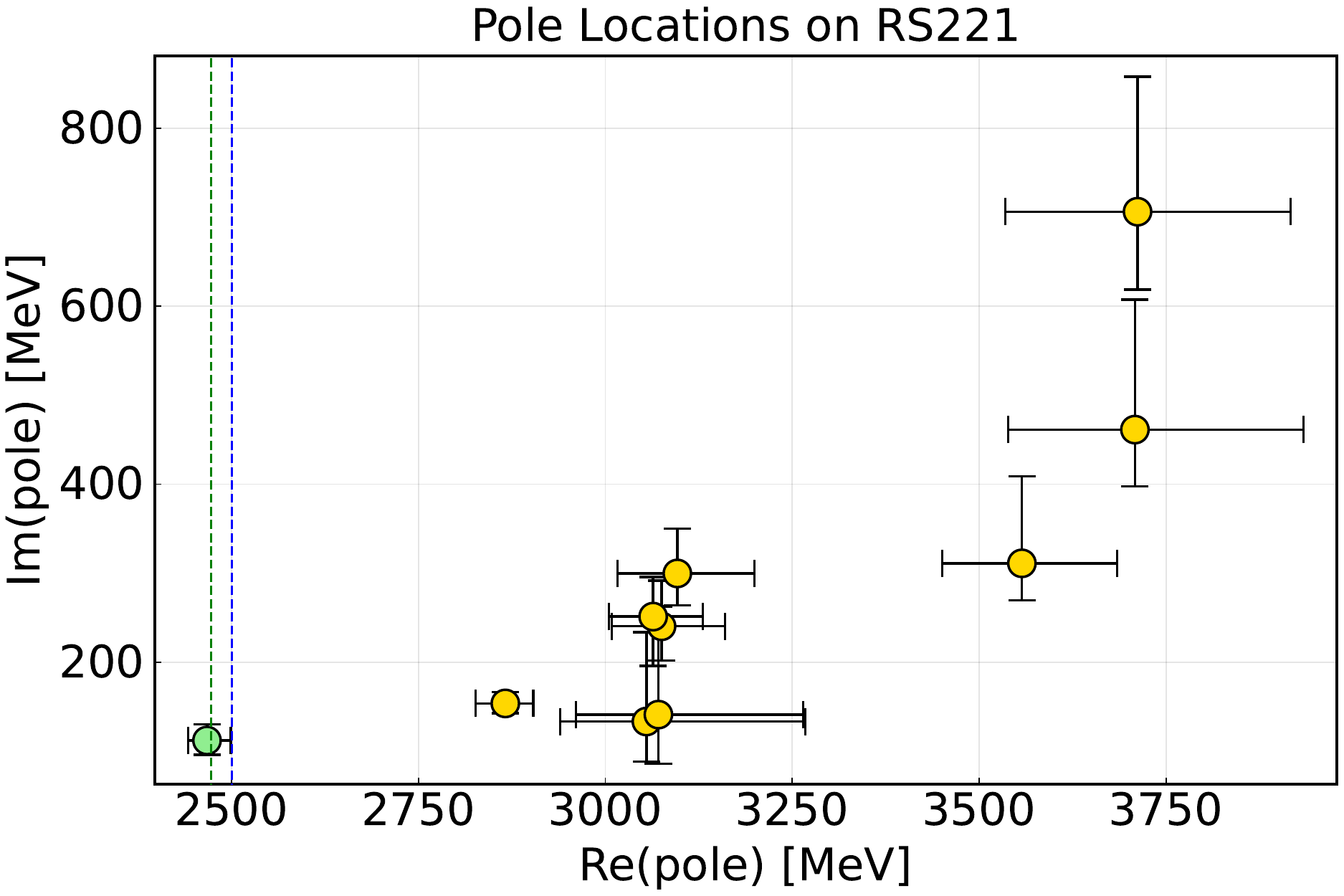}
    \caption{The location of poles on sheet RS221 on the complex energy plane. The 
    $x$-axis and $y$-axis show the real and imaginary part of energy, respectively. The 
    poles from the amplitude parametrizations employed in Ref.~\cite{Moir:2016srx} are 
    shown in yellow. The pole from the UChPT amplitude~\cite{Liu:2012zya} is shown in green~\cite{Albaladejo:2016lbb}.
    The vertical green and blue dashed lines represent the $D\eta$ and $D_s\Bar{K}$ thresholds, respectively.
    The error bars show the $1\sigma$\ statistical uncertainty.}
    \label{fig:plocRS221}
\end{figure}

\begin{table*}
    \centering
    \caption{The pole locations from amplitude parametrizations of Ref.~\cite{Moir:2016srx}, in units of MeV. Empty slots
    denote that a pole was not found within the search range for the particular parametrization on the 
    corresponding sheet. 
    In the last line the results for the UChPT amplitude employed in Ref.~\cite{Albaladejo:2016lbb}
    are given for comparison.}
\begin{tabular*}{\textwidth}{@{\extracolsep{\fill}}ccccc@{}} 
            \hline\noalign{\smallskip}
        Amplitudes & RS111 & RS211 & RS221 & RS222  \\ 
            \noalign{\smallskip}\hline\noalign{\smallskip}
        Amplitude 1 & $2275.92 $ & $2720_{-89}^{+150}\, - i\,198_{-70}^{+37} $& $3060_{-120}^{+210}\, - i\,133_{-45}^{+100}$  & $3030_{-130}^{+260}\, - i\,430_{-120}^{+210}$ \   \\ 
        Amplitude 2 & $2275.92$ & $2717_{-51}^{+99}\, - i\,204_{-47}^{+45} $ & $3070_{-110}^{+190}\, - i\,141_{-55}^{+120} $ & $3080_{-120}^{+230}\, - i\,420_{-120}^{+200}$ \ \\ 
        Amplitude 3 & $2275.92$ &  & $3710_{-180}^{+210}\, - i\,706_{-87}^{+150} $ & $3880_{-200}^{+210}\, - i\,1153_{-79}^{+110}$ \ \\ 
        Amplitude 4 & $2275.94 $&  & $3710_{-170}^{+230}\, - i\,461_{-64}^{+150} $ & $3840_{-180}^{+260}\, - i\,763_{-58}^{+130} $\   \\ 
        Amplitude 5 & $2276.04 $& $2789_{-54}^{+590}\, - i\,27_{-27}^{+210} $ & $3560_{-110}^{+130}\, - i\,311_{-42}^{+98} $ & $3680_{-120}^{+140}\, - i\,601_{-40}^{+91}$ \  \\ 
        Amplitude 6 & $2275.70 $& $2618_{-50}^{+64}\, - i\,240_{-35}^{+56}$  & $3075_{-67}^{+85}\, - i\,240_{-38}^{+51} $ & $3162_{-82}^{+100}\, - i\,349_{-45}^{+64}$ \   \\ 
        Amplitude 7 & $2275.98 $& $2652_{-53}^{+70}\, - i\,291_{-41}^{+57}$   & $3096_{-80}^{+100}\, - i\,300_{-36}^{+51} $ & $3180_{-91}^{+120}\, - i\,410_{-55}^{+70}$ \  \\ 
        Amplitude 8 & $2275.70 $& $2621_{-67}^{+90}\, - i\,242_{-31}^{+34}$  &$ 3064_{-59}^{+66}\, - i\,251_{-55}^{+45}  $ & $3141_{-74}^{+80}\, - i\,318_{-48}^{+96} $\   \\ 
        Amplitude 9 & $2275.92 $& $2673_{-25}^{+28}\, - i\,182_{-26}^{+23} $ & $ 2866_{-40}^{+38}\, - i\,154_{-11}^{+13} $ & $2909_{-38}^{+38}\, - i\,274_{-22}^{+26}$ \ \\ 
        \hline
        \hline
        UChPT  & $2263^{+8}_{-14} $& $2633_{-43}^{+79} \, - i\,114_{-12}^{+11} $ & $ 2467_{-25}^{+32}\, - i\,113_{-16}^{+18} $ & $3000_{-110}^{+290}\, - i\,93_{-15}^{+21}$ \ \\ 
            \noalign{\smallskip}\hline
        \end{tabular*}
    \label{tab:poles}
\end{table*}
Graphically the poles on RS221 are displayed in Fig.~\ref{fig:plocRS221}.
In the following we focus the discussion on this sheet, since this is the one 
where the UChPT amplitude has its most prominent higher $D_0^*$ pole at physical~\cite{Du:2017zvv} 
as well as the unphysical meson masses 
employed in the lattice study~\cite{Albaladejo:2016lbb}. The plots of the pole locations of the higher pole for the
different parametrizations
on the 
other Riemann sheets
that connect closely to the physical axis (RS211 and RS222) are shown in the Appendix. Table~\ref{tab:thresholds} gives the location of the 
corresponding two particle thresholds.

\begin{table}
    \centering
     \caption{The two particle thresholds in MeV for the pion mass of 391~\mev.}
    \begin{tabular}{cc}
    \hline\noalign{\smallskip}
         Threshold &  Threshold [MeV]\\
         \noalign{\smallskip}\hline\noalign{\smallskip}
         $D\pi$ & $2276.49$ \\
         $D\eta$ & $2472.46$ \\
         $D_s\Bar{K}$ & $2500.51$ \\ 
         \noalign{\smallskip}\hline
    \end{tabular}
    \label{tab:thresholds}
\end{table}

Figure~\ref{fig:plocRS221} and Fig.~\ref{fig:plocRS211RS222}  in the Appendix and Table~\ref{tab:poles} clearly show two important features
of the poles extracted from different parametrizations:
$(i)$ There is a significant correlation between
real part and imaginary part of the poles, and the location of
the pole extracted from the UChPT analysis is in line with
that correlation. $(ii)$ All poles are located on hidden 
sheets, which are the sheets that are not directly connected to the physical sheet. 
For example, the RS221 poles are well above the $D_s\Bar{K}$\ threshold. 
Thus they are all shielded by the RS222 sheet and their effect on the amplitude 
can hardly be seen above the $D_s\bar{K}$\ threshold.
As we discuss in the following, both features together guide one to an
understanding that there indeed needs to be a second pole 
in an amplitude that describes the lattice data and that it 
is natural that the original analysis performed on the lattice
data lead to badly constrained pole locations. 
The mechanism underlying this is that
the distance from the threshold is overcome by an enhanced residue. 
This mechanism, also reported e.g. for the case of the $f_0(980)$ and $a_0(980)$,
was observed before as a general feature of Flatt\'e amplitudes~\cite{Baru:2004xg}.

\subsection{Residues and Threshold distance}
\label{subsec:DxR}
A resonance is characterized  by the pole
location, traditionally parametrized as
\begin{equation}
    \sqrt{s_p} = M-i\Gamma/2 .
    \label{eq:MGamdef}
\end{equation}
Please note that the parameters $M$ and $\Gamma$, derived from the pole location,
agree to those found e.g. in the BW fits only for narrow,
isolated resonances --- for details see the review on
resonances in Ref.~\cite{Workman:2022ynf}.
Equally fundamental resonance properties
are provided by the pole residues. A pole-residue
quantifies the couplings of the resonance to the various channels. 
The residues of a pole located at $s=s_p$ are defined as 
\begin{equation}
     R_{ij} = \lim_{s\to s_p}(s-s_p){T}_{ij}(s).
     \label{eq:resdef}
\end{equation}
The residues can be easily obtained using the L'H\^opital rule to compute the limit: 
\begin{equation}
    R_{ij} = \left(\frac{d}{ds}{T}_{ij}^{-1} \right)^{-1} _{s=s_p}.
\end{equation}

Since the residues factorize according to $R_{ij}^2=R_{ii}^{}R_{jj}^{}$
one can define an effective coupling via
\begin{equation}
    g^r_i = R_{ij}/\sqrt{R_{jj}} ,
\end{equation}
which has dimension [mass].
The index $r$ is meant to distinguish the residues from the parameters 
$g_i$ that appear in
the $K$-matrix in Eq.~(\ref{eq:Kdef}).
The couplings $g^r_i$ characterize the transition strengths of the resonance to the 
channel. Those residues can also be extracted from production reactions and are
independent of how the resonance was 
produced.

Since the poles of interest here are hidden, their effect on the physical axis is
visible only at the thresholds irrespective of their exact
pole locations. Moreover, the visible effect in the
amplitude on the physical axis from
a pole on a hidden sheet 
close to the threshold with a small residue is in fact hardly distinguishable 
from a faraway pole with a large residue. 
We regard this ambiguity as the most natural explanation 
for the large spread in the pole locations found in
the analysis of the lattice data reported 
above.

To test this hypothesis,
we now study the strengths of the residues as functions of 
the distance of the poles to the threshold.
Clearly, there is some ambiguity in how to quantify the distance
from the threshold. Since the channel couplings also
drive the size of the imaginary part of the pole location
and we want to avoid counting the effect of those couplings twice,
we choose instead of $\sqrt{(M - M_{\rm thr.})^2 + (\Gamma/2)^2 }$,
which might appear more natural on the first glance,
\begin{align}
    {\rm Dist} = M - M_{\rm thr.}
    \label{eqn:pole distance}
\end{align}
as a measure for the distance of the pole to the threshold.
The pole mass $M$ that appears above was introduced in Eq.~(\ref{eq:MGamdef})
and $M_{\rm thr.}$ denotes the threshold location relevant for the
given sheet, e.g. in case of RS221 we have $M_{\rm thr.}=M_K+M_{D_s}$.

\begin{figure*}
    \centering
    \includegraphics[width=0.33\textwidth]{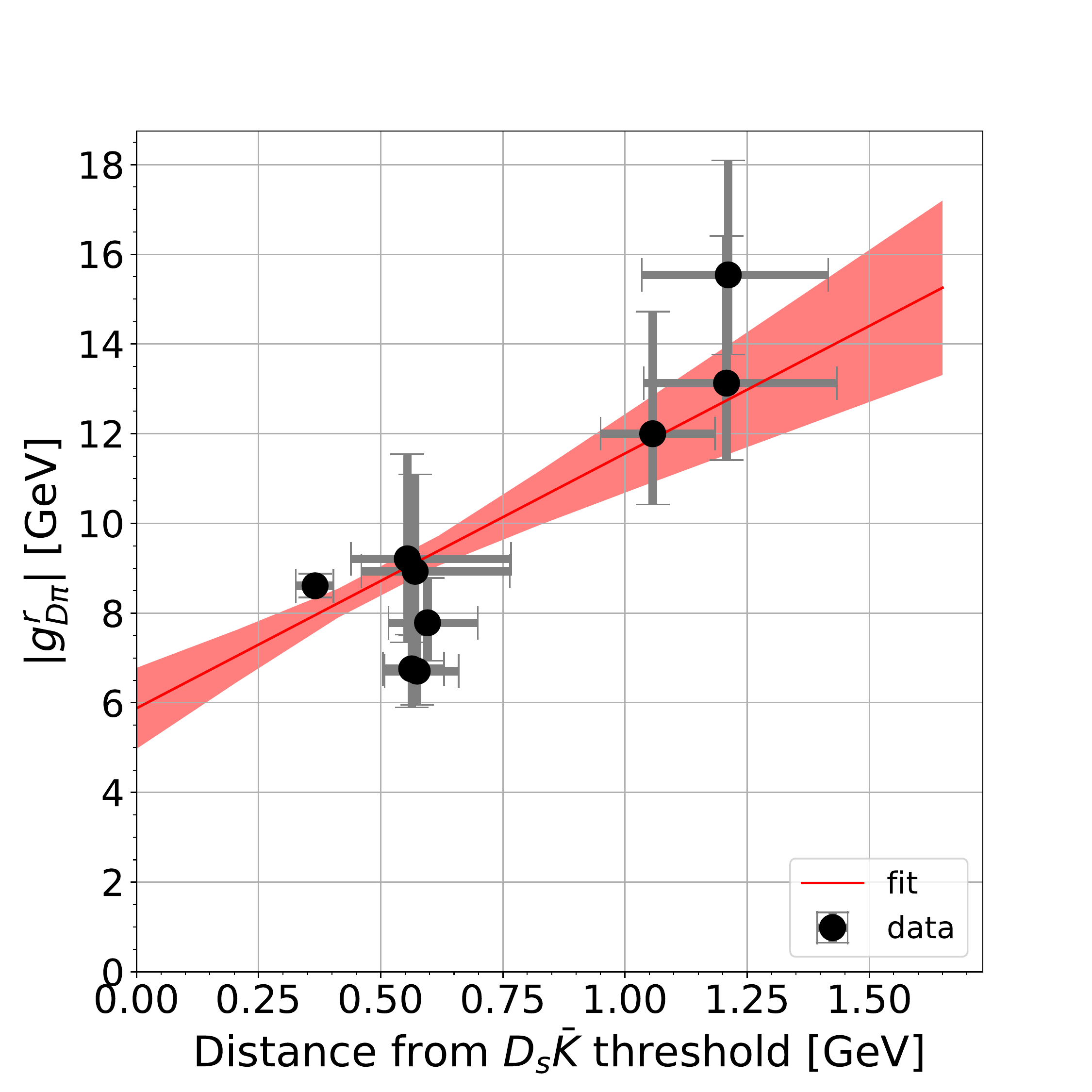}
    \includegraphics[width=0.33\textwidth]{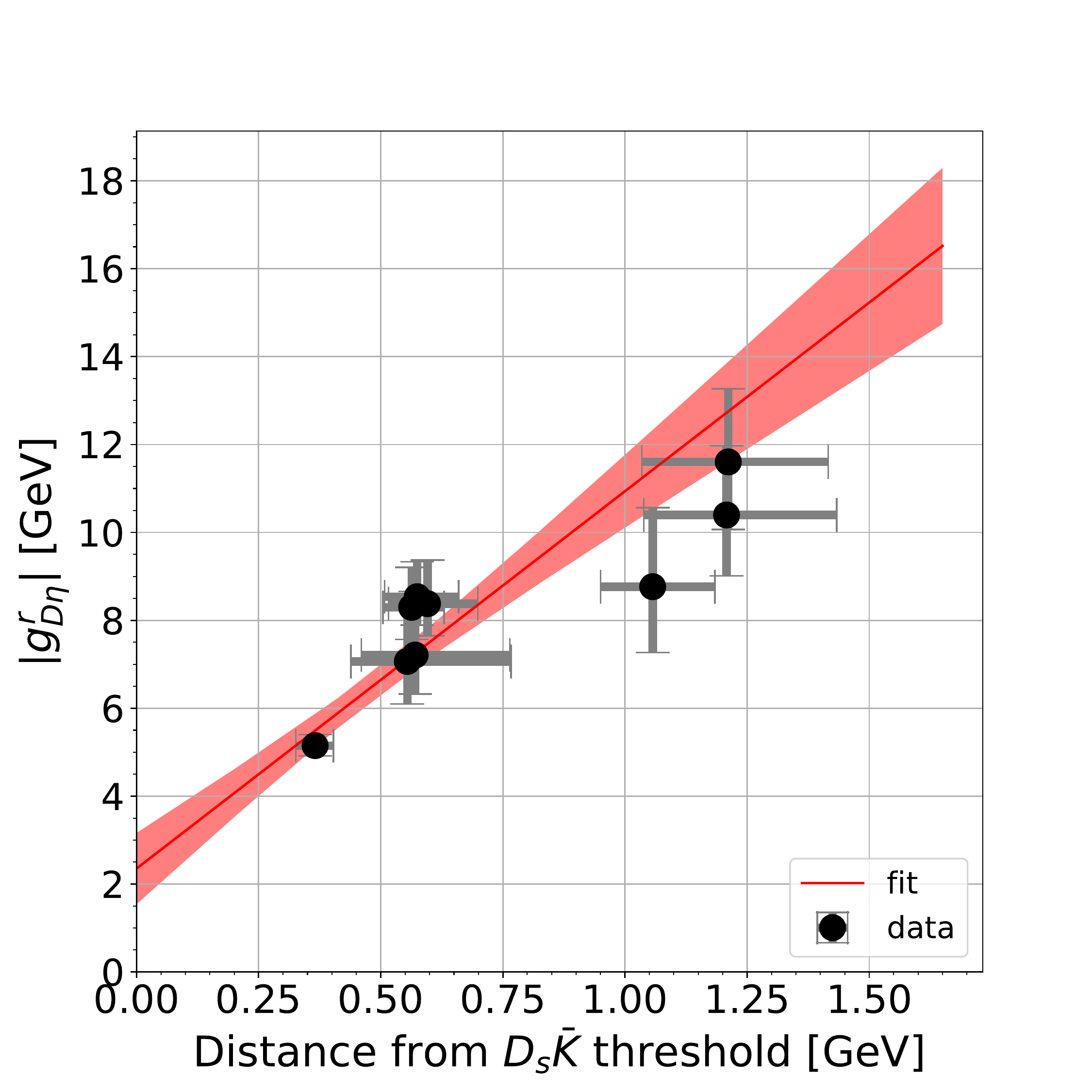}
    \includegraphics[width=0.33\textwidth]{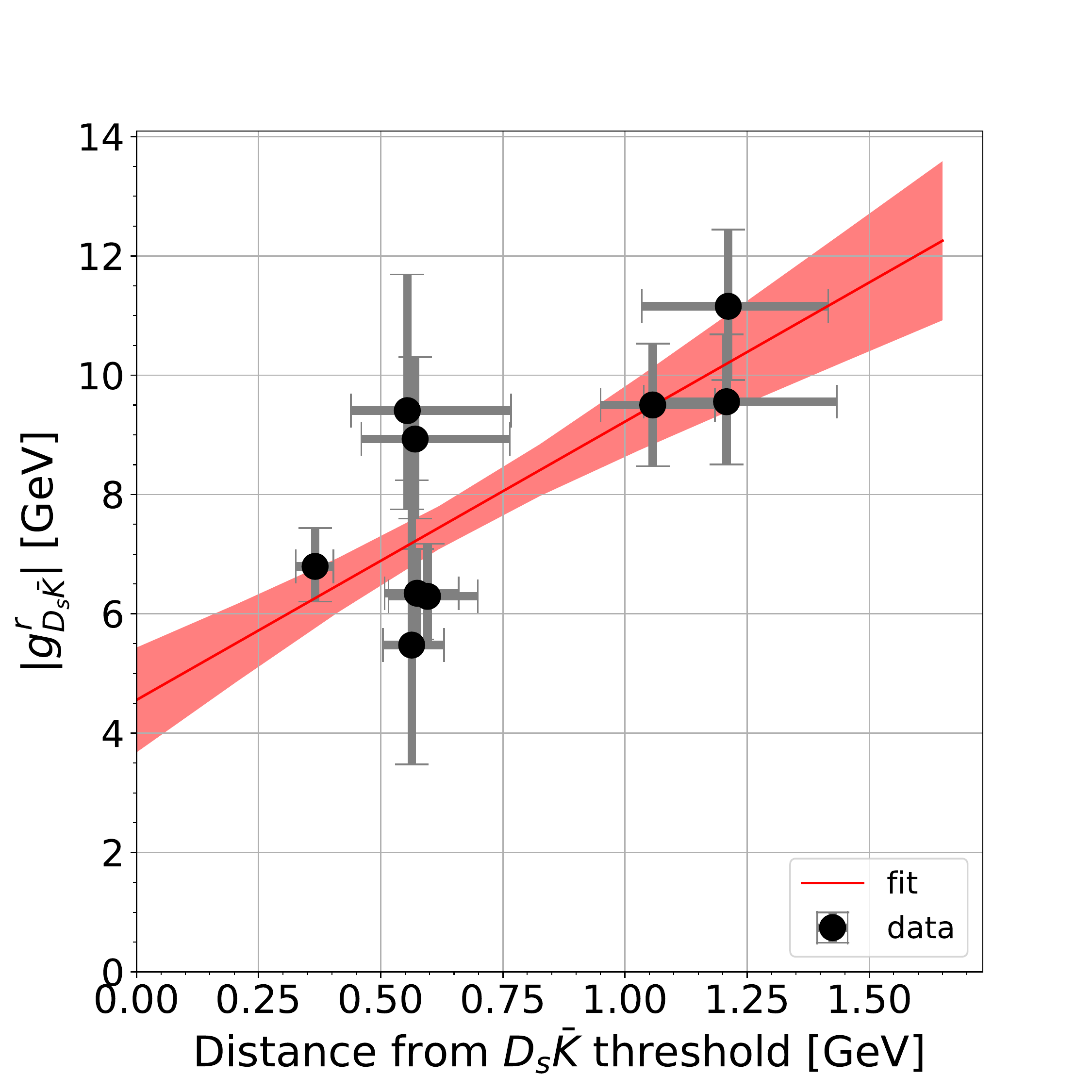}
    \caption{The distance of the real part of the pole on sheet RS221 from the $D_s\Bar{K}$\ 
    threshold versus the effective coupling of the pole to the 
    $D\pi$ channel (left), $D\eta$ channel (middle) and $D_s\Bar{K}$ channel (right). The red line shows the 
    straight line fit. The red band encloses the $1\sigma$\ uncertainty of the fit. 
    }
    \label{fig:RS221Res}
\end{figure*}

Table~\ref{tab:dxr} shows the distance of the RS221 pole from the 
$D_s\Bar{K}$\ threshold along with the square root of the absolute value of 
the residue to the three channels. The graphical representation for the
$D\pi\text-D\pi$ channel is shown in Fig.~\ref{fig:RS221Res} (left).
A straight line is fit to the data to extrapolate the values at the 
threshold. The fitting was done 
using the MINUIT algorithm \cite{James:1975dr} from the \textsf{iminuit} 
interface ~\cite{hans_dembinski_2022_7115916,iminuit.jl}. The uncertainties of the fit parameters 
quoted are from the MIGRAD routine of MINUIT. The $1\sigma$\ statistical 
uncertainty of the fitted line was calculated using the
bootstrap technique. From the straight line fit 
the $y$-intercept was found to be at $(5.8\pm{0.9})$~\gev. 
The graphical representation for corresponding distances and residues for the $D\eta\text-D\eta$ and the
$D_s\Bar{K}\text-D_s\Bar{K}$ channel 
is given in Fig.~\ref{fig:RS221Res} (middle) and  Fig.~\ref{fig:RS221Res} (right).
The fit to the $D\eta$ ($D_s\Bar{K}$) residues provides an
intercept of $(2.4\pm{0.8})$~\gev\, ($(4.6\pm{0.9})$~\gev). 
The corresponding results for the poles on sheets RS211 
and RS222 are shown in the Appendix.

While the linear fits shown do not provide an excellent representation 
of the extracted data for the different channels, they
illustrate nicely that there is indeed a significant correlation
between the distance of the poles from the threshold
and the residues. In addition,
the effect of the poles at threshold, encoded in the 
$y$ intercepts deduced from the fits, is rather well constrained by 
the fits.
We interpret this observation such that the lattice data 
require not only one bound state pole but also  a higher pole
as was also found in the various studies employing UChPT~\cite{Kolomeitsev:2003ac,Guo:2006fu,Guo:2006rp,Guo:2009ct,Albaladejo:2016lbb,Du:2017zvv,Lutz:2022enz}.

An interesting question is, if it is possible to come up with a 
parametrization to be used in the $K$-matrix fit that constrains
better the pole location of the higher pole, with inputs of approximate
symmetries of QCD. This will be the focus of the next section.

\begin{table*}
    \centering
    \caption{Distance of the RS221 pole from the $D_s\Bar{K}$ threshold and the square root of its residue to the respective channels in MeV  
    for the amplitude parametrizations obtained in Ref.~\cite{Moir:2016srx}.
    In the last line the results for the UChPT amplitude employed in Ref.~\cite{Albaladejo:2016lbb}
    are given for comparison.}
\begin{tabular*}{\textwidth}{@{\extracolsep{\fill}}ccccc@{}} 
            \hline\noalign{\smallskip}
        Amplitudes & Dist. from $D_s\Bar{K}$ thr. [\mev] & $|\gdp|$ [\gev] & $|\gdeta|$ [\gev] & $|\gdsk|$ [\gev] \\
          \noalign{\smallskip}\hline\noalign{\smallskip}
          Amplitude 1 &  $554_{-115}^{+212}$ & $9_{-2}^{+2}$ & $7_{-1}^{+1}$ & $9_{-2}^{+2}$ \\
          Amplitude 2 & $570_{-110}^{+194}$ & $9_{-1}^{+2}$ & $7_{-1}^{+2}$ & $9_{-1}^{+1}$ \\
          Amplitude 3 & $1211_{-177}^{+205}$ & $16_{-2}^{+3}$ & $12_{-2}^{+2}$ & $11_{-1}^{+1}$ \\
          Amplitude 4 & $1208_{-170}^{+226}$ & $13_{-2}^{+3}$ & $10_{-1}^{+2}$ & $10_{-1}^{+1}$ \\
          Amplitude 5 & $1057_{-107}^{+127}$ & $12_{-2}^{+3}$ & $9_{-2}^{+2}$ & $10_{-1}^{+1}$ \\
          Amplitude 6 & $596_{-80}^{+103}$ & $8_{-1}^{+1}$ & $8_{-1}^{+1}$ & $6.3_{-0.8}^{+0.9}$ \\
          Amplitude 7 & $575_{-67}^{+85}$ & $7_{-1}^{+1}$ & $8.5_{-0.6}^{+0.8}$ & $6.3_{-0.8}^{+0.7}$ \\
          Amplitude 8 &  $563_{-59}^{+66}$ & $6.8_{-0.9}^{+0.8}$ & $8.3_{-0.7}^{+0.9}$ & $6_{-2.}^{+3}$ \\
          Amplitude 9 & $366_{-40}^{+38}$ & $8.6_{-0.3}^{+0.3}$ & $5.1_{-0.2}^{+0.2}$ & $6.8_{-0.6}^{+0.6}$ \\
          \hline
          \hline
          UChPT & $-34_{-25}^{+32}$ & $5.2_{-0.4}^{+0.6}$ & $6.7_{-0.4}^{+0.6}$ & $13_{-1}^{+2}$ \\
            \noalign{\smallskip}\hline
        \end{tabular*}
    \label{tab:dxr}
\end{table*} 

\section{SU(3) symmetry}
\label{sec:SU3sym}
The parametrization dependence of the higher pole location calls for a stronger constrained amplitude. In the following we present a prescription of the $K$-matrix consistent
with the SU(3) flavor symmetry. In the resulting scattering matrix SU(3) breaking 
comes only from the Chew-Mandelstam
functions introduced in Eq.~(\ref{eq:chew}).
Clearly, for the physical pion mass and low energies such a treatment is not justified, since
the leading order chiral interaction scales with the energies of pion
and kaons for $D \pi$ and $D_s\bar{K}$ scattering, respectively, which induces a 
sizeable SU(3) flavor breaking into the scattering potential. However, in this study
we work at higher pion masses which leads to a much smaller pion-kaon mass difference.
Moreover,
we are mainly interested in the higher mass range, where the second pole 
is located. Under such circumstances the leading SU(3) breaking effect is induced by the 
loop functions which bring the cut structure to the amplitudes. 

\begin{figure}
    \centering
    \resizebox{\columnwidth}{!}{%
    \begin{tikzpicture}[%
        dot/.style={circle, fill, inner sep=2pt}
        ]   
    
    \node[dot] (11) {}; 
    \node[dot, right=of 11] (12) {}; 
    \draw (11)--(12);

    \node[dot, below left=1cm and 5mm of 11 ](21){};
    \node[dot, right=of 21](22){};
    \draw[inner sep=0pt](22) circle (0.2) ;
    \node[dot, right=of 22](23){};
    
    \draw (21)--(23);
    \draw (11)--(21);
    \draw (11)--(22);
    \draw (12)--(23);
    \draw (12)--(22);

    \node[dot, right=1.5cm of 23](24){};
    \node[dot, right=of 24](25){};
    \node[dot, right=of 25](26){};

    \draw (24)--(26);

    \node[dot, right=1.5cm of 26](27){};
    
    \node[dot, below left=1cm and 5mm of 21 ](31){};
    \node[dot, right=of 31](32){};
    \draw[inner sep=0pt](32) circle (0.2) ;
    \node[dot, right=of 32](33){};
    \draw[inner sep=0pt](33) circle (0.2) ;
    \node[dot, right=of 33](34){};

    \draw (31)--(34);
    \draw (21)--(31);
    \draw (21)--(32);
    \draw (22)--(33);
    \draw (22)--(32);
    \draw (34)--(23);
    \draw (23)--(33);

    \node[dot, right=1.5cm of 34](35){};
    \node[dot, right=of 35] (36) {}; 
    \draw (35)--(36);

    \draw (24)--(35);
    \draw (25)--(35);
    \draw (25)--(36);
    \draw(26)--(36);

    \node[dot, right=1.5cm of 36](37){};
    \node[dot, right=of 37] (38) {}; 
    \draw (37)--(38);
    \draw (37)--(27);
    \draw (27)--(38);

    \node[dot, below right=1cm and 5mm of 31 ](41){};
    \node[dot, right=of 41](42){};
    \node[dot, right=of 42](43){};

    \draw (41)--(43);
    \draw (31)--(41);
    \draw (32)--(41);
    \draw (32)--(42);
    \draw (33)--(42);
    \draw (34)--(43);
    \draw (33)--(43);

    \node[dot, below right=1 and 5mm of 35 ](44){};

    \draw (35)--(44);
    \draw (36)--(44);

    \node[left= 1.5cm of 11] (1) {$S=2$};
    \node at (1|-21) {$S=1$};
    \node at (1|-31) {$S=0$};
    \node at (1|-41) {$S=-1$};

    \end{tikzpicture}}
\caption{Weight diagrams of the $[\overline{15}], [6]$\ and $[\Bar{3}]$\ representations.}
    \label{fig:multiplet}
\end{figure}
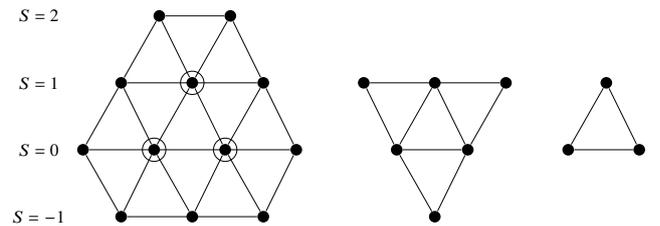

The flavor structure of the  $D\Phi$\ interaction can be written as a direct product of an 
anti-triplet for the charmed mesons and an octet for the light pseudoscalar mesons. The direct product can be decomposed into a direct sum of the $[\Bar{3}]$, 
$[6]$\ and $[\overline{15}]$\ irreducible representations. Figure~\ref{fig:multiplet} shows the multiplet 
structure.  

The SU(3) flavor basis and the isospin-symmetric particle basis are related via
\begin{equation}
    \begin{pmatrix}
    \lvert [\Bar{3}] \rangle \\ 
    \lvert [6] \rangle \\
    \lvert [\overline{15}] \rangle 
    \end{pmatrix}
    = U
    \begin{pmatrix}
    \lvert D\pi \rangle \\ 
    \lvert D\eta \rangle \\
    \lvert D_{s}\Bar{K} \rangle 
    \end{pmatrix} ,
\end{equation}    
where
\begin{equation}
    U=
    \begin{pmatrix}
     -3/4 & -1/4 & -\sqrt{3/8}
   \\
   \sqrt{3/8} & -\sqrt{3/8} & -1/2
   \\
    1/4 & 3/4 & -\sqrt{3/8}
    \\
    \end{pmatrix}  .
    \label{eqn:basisrelation}
\end{equation}
From the rotation matrix $U$, we may read off the following 
expressions for the SU(3) symmetric coupling structure of the $D\pi$, $D\eta$\ and 
$D_s\Bar{K}$~$(I =\frac{1}{2}, S = 0)$\ coupled-channel system:
\begin{align}
     C_{\Bar{3}} &= \begin{pmatrix}
     -3/4 \\ -1/4 \\ -\sqrt{3/8}
   \end{pmatrix}  \begin{pmatrix}
     -3/4 & -1/4 & -\sqrt{3/8}
   \end{pmatrix} \nonumber \\
   &= \frac{3}{8} \begin{pmatrix}
     3/2 & 1/2 & \sqrt{3/2}
   \\
   1/2 & 1/6 & \sqrt{1/6}
   \\
    \sqrt{3/2} & \sqrt{1/6} & 1
    \end{pmatrix}~,
    \end{align}
    \begin{align}
     C_6 &=  \begin{pmatrix}
     \sqrt{3/8} \\ -\sqrt{3/8} \\ -1/2
     \end{pmatrix}  \begin{pmatrix}
     \sqrt{3/8} & -\sqrt{3/8} & -1/2
     \end{pmatrix} \nonumber \\
    &= \frac12 \begin{pmatrix}
     3/4 & -3/4 & -\sqrt{3/8}
   \\
   -3/4 & 3/4 & \sqrt{3/8}
   \\
    -\sqrt{3/8} & \sqrt{3/8} & 1/2
    \end{pmatrix}~,
    \end{align}
    \begin{align}
     C_{\overline{15}} &= \begin{pmatrix}
     1/4 \\ 3/4 \\ -\sqrt{3/8}
     \end{pmatrix}\begin{pmatrix}
     1/4 & 3/4 & -\sqrt{3/8}
     \end{pmatrix} \nonumber \\
     &= \frac{3}{8}   \begin{pmatrix}
     1/6 & 1/2 & -\sqrt{1/6}
   \\
   1/2 & 3/2 & -\sqrt{3/2}
   \\
    -\sqrt{1/6} & -\sqrt{3/2} & 1
    \end{pmatrix}~.
    \label{eqn:factormatrices}
\end{align}    
The form of the $K$-matrix assuming the existence of two bare poles, in contrast to the one used in Ref.~\cite{Moir:2016srx} which contains only one bare pole, reads
\begin{align}
 K = \left(\frac{g_{\Bar{3}}^2}{m_{\Bar{3}}^2-s}{+} c_{\Bar{3}}\right)C_{\Bar{3}} + \left(\frac{g_{6}^2}{m_{6}^2-s}{+} c_{6}\right)C_{6} + 
 c_{\overline{15}}\,C_{\overline{15}}.
 \label{eqn:su3k}
\end{align}
Here the two bare poles are assumed to be in the two SU(3) multiplets with $S$-wave attractions from the leading order chiral dynamics~\cite{Albaladejo:2016lbb}.
We have seven free parameters in total, $g_{\alpha}$, $c_{\alpha}$ and $m_\alpha$. 
The overall 
factors in Eq.~\eqref{eqn:factormatrices} are absorbed into the parameters of $g_{\alpha}$, $c_{\alpha}$.
If there was no SU(3) constraint, a $K$-matrix with the same number of bare poles would contain 3 more parameters (a constant $K$ matrix is symmetric and thus contains 6 parameters, instead of 3 $c_\alpha$'s here).

With Eqs.~\eqref{eq:Tmat} and \eqref{eq:chew}, the $T$-matrix $T_K(s)$ can be calculated in the same way as in Sect. \ref{sec:lanalysis}.
The subtraction point for the Chew-Mandelstam function of the fits
is chosen identical to the parameter $m_{\Bar{3}}$.

\section{Fitting to Lattice Energy Levels}
\label{sec:latfits}
Here we employ the  flavor SU(3) constrained 
$K$-matrix
to fit the lattice energy levels in the $D\pi$ c.m. frame obtained in 
Ref.~\cite{Moir:2016srx}. 
To this end, we need to relate the $T$-matrix defined with the $K$-matrix in the continuum system and the energy levels in the finite volume system.
In this study, we employ the scheme based on the effective field theory framework developed in Ref.~\cite{Doring:2011vk}. Below we briefly summarize this scheme.

With the Lippmann-Schwinger equation, the $T$-matrix in the continuum can  be written as
\begin{align}
    T(s) = \frac{1}{V^{-1}(s) - G(s)},
    \label{eqn:Tmatlipp}
\end{align}
where $V(s)$ is the interaction matrix and $G(s)$ is the 
diagonal matrix of the scalar two-meson loop functions~\cite{Oller:2000fj}. With the momentum cut-off
regularisation $G(s)$ is given by
\begin{align} \nonumber
    G_{ii}(s) &= \int\displaylimits^{\left|\Vec{q}\right|<q_{\rm max}} 
    \frac{d^3\Vec{q}}{(2\pi)^3}\frac{1}{2\omega_1^{(i)}(\left|\Vec{q}\right|)\omega_2^{(i)}(\left|\Vec{q}\right|)} \\ &
    \hspace{1cm} \times \frac{\omega_1^{(i)}(\left|\Vec{q}\right|)+
    \omega_2^{(i)}(\left|\Vec{q}\right|)}{s-(\omega_1^{(i)}(\left|\Vec{q}\right|)+\omega_2^{(i)}(\left|\Vec{q}\right|)^2},
    \label{G}\\
 \omega^{(i)}_{1,2}(\Vec{q}) & = \sqrt{m^{2(i)}_{1,2} + \Vec{q}^2},
\end{align}
where $q_{\rm max}$\ is the cut-off momentum and 
$m^{(i)}_{1,2}$\ are the masses of the two particles in channel $i$.
Similarly to Eq.~\eqref{eqn:Tmatlipp}, the $T$-matrix in a finite volume system $\Tilde{T}$ satisfies 
\begin{align}
    \tilde{T}(s) = \frac{1}{\tilde{V}^{-1}(s) - \tilde{G}(s)},
    \label{eqn:Tmatlipp_finite}
\end{align}
where $\tilde{V}(s)$ and $\tilde{G}(s)$ are the interaction and loop function in the finite volume system, respectively.
With the spatial extension $L$ of the cubic box, $\tilde{G}(s)$ is given as 
\begin{align}\nonumber
    \Tilde{G}_{ii}(s) =&\, \frac{1}{L^3} \sum_{\Vec{q}}^{\left|\Vec{q}\right|<q_{\rm max}} 
    \frac{1}{2\omega_1^{(i)}(|\Vec{q}|)\omega_2^{(i)}(|\Vec{q}|)} \\ & \times
    \frac{\omega_1^{(i)}(|\Vec{q}|)+\omega_2^{(i)}(|\Vec{q}|)}{s-(\omega_1^{(i)}(|\Vec{q}|)+
    \omega_2^{(i)}(|\Vec{q}|))^2}, 
 \end{align}
with 
\begin{align}
    \vec{q} = \frac{2\pi}{L}\vec{n},\, \vec{n}\in \mathbb{Z}^3.
\end{align}
Since $V(s)$ is equal to $\tilde{V}(s)$ up to exponentially suppressed corrections,
the $T$-matrix in the finite volume system $\Tilde{T}$ is related to 
the $T$-matrix in the infinite volume $T$\  by
\begin{align}
    \label{eqn:Tfinite}
    \Tilde{T}(s) &= \frac{1}{T^{-1}(s) - \Delta G(s)},
\end{align}
with
\begin{align}
    \Delta G_{ii} &= \Tilde{G}_{ii}(s) - G_{ii}(s).
\end{align} 
The lattice energy levels, which we need to fit, correspond to the zeroes of the determinant of $\tilde{T}^{-1}$
provided in Eq.~(\ref{eqn:Tfinite}).

\subsection{Results for the fits employing SU(3) constraints}
\label{subsec:prelimfit}

We performed four different fits to the rest frame lattice energy levels in Ref.~\cite{Moir:2016srx}
using the MINUIT algorithm~\cite{James:1975dr} with the Julia interface to the \textsf{iminuit} package~\cite{hans_dembinski_2022_7115916,iminuit.jl}:
\begin{itemize}
\item in Fit 1\_4L all the parameters in Eq.~(\ref{eqn:su3k}) are included;
\item in Fit 2\_4L we fix $c_{\bar 3} = 0 $ and $c_{6} = 0$;
\item in Fit 3\_4L we fix $c_{6} = 0$;
\item in Fit 4\_4L we fix $g_{6} = 0$ to omit the explicit pole term of $[6]$ (and thus $m_6$ is absent).
\end{itemize}

\begin{figure}
    \centering
    \includegraphics[scale=0.44]{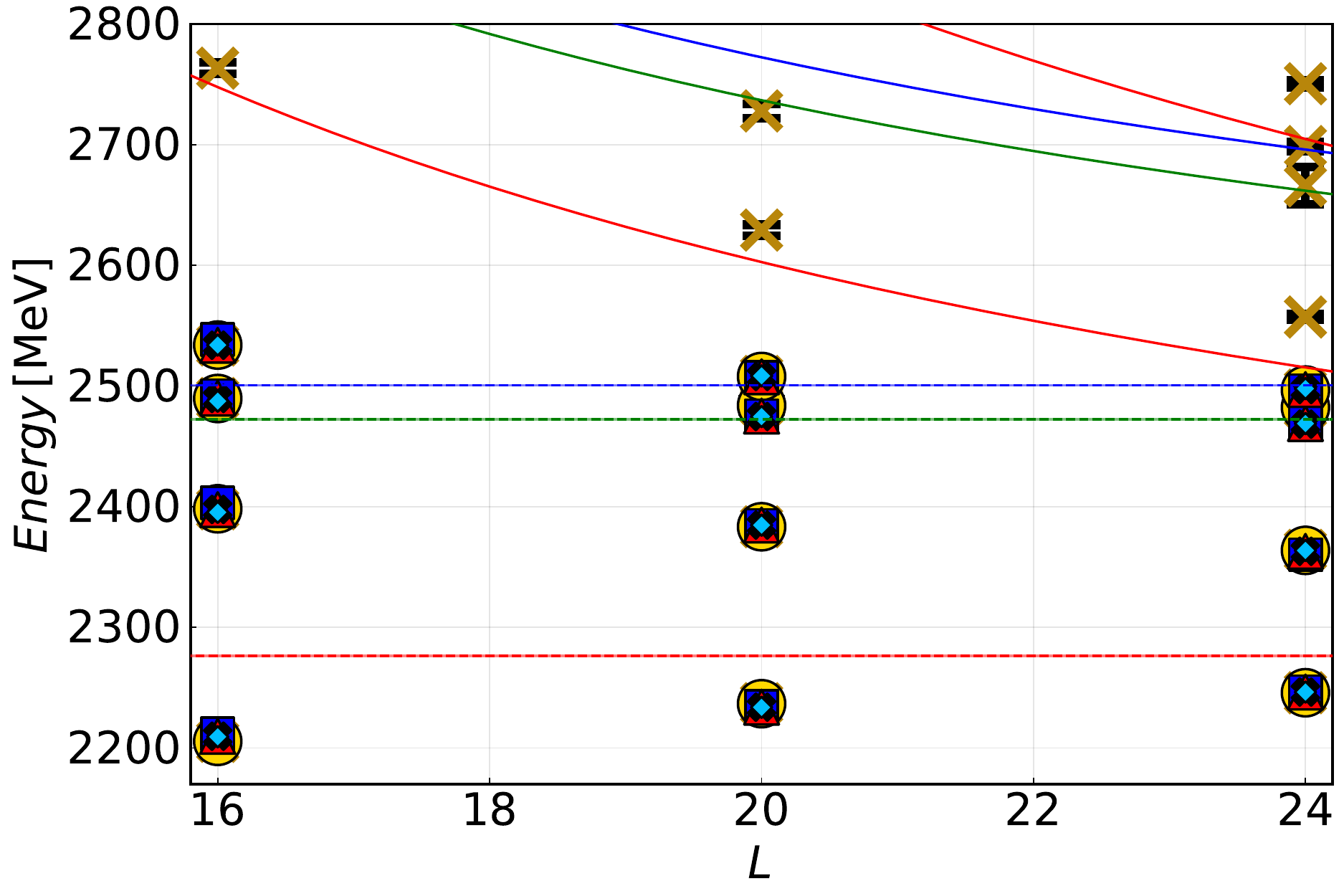}
    \caption{Comparison of the energy levels from data and the fits. The energy levels from Ref.~\cite{Moir:2016srx} are shown in yellow. The energy levels used as input in the fits are shown as circles and those not used is shown as crosses.
    The energy levels from Fit~1\_4L, Fit~2\_4L, Fit~3\_4L and Fit~4\_4L are shown in red, blue, black and light blue, respectively.
    The solid(dashed) red, green and blue lines show the $D\pi$, $D\eta$ and
    $D_s\Bar{K}$ non-interacting energy levels(thresholds) respectively.    }
    \label{fig:elvlscomb}
\end{figure}

\begin{figure}
    \centering
    \includegraphics[scale=0.44]{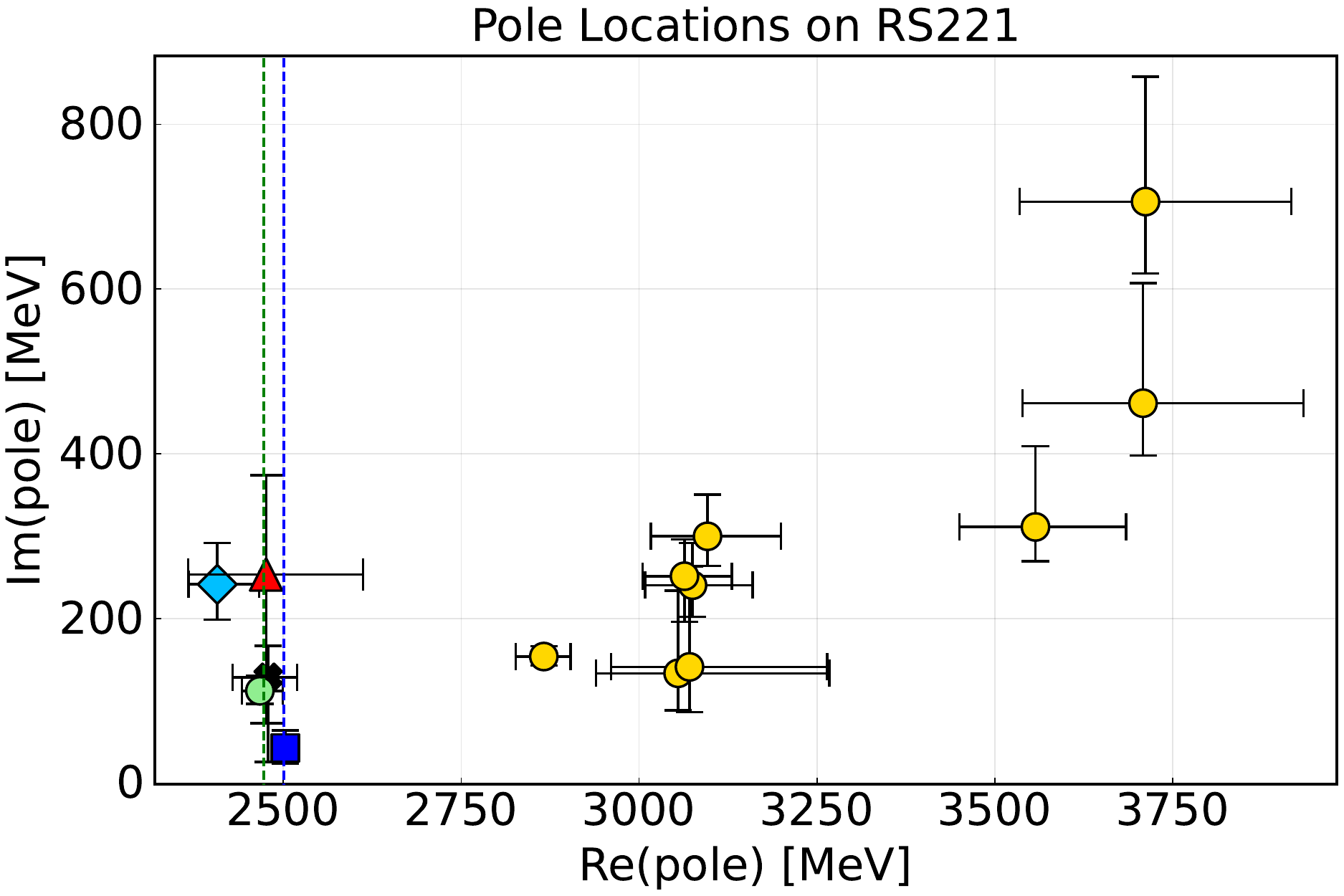}
    \caption{Locations of the RS221 poles from the different fits
    based on Eq.~(\ref{eqn:su3k}), together with the pole reported
    in UChPT amplitude~\cite{Albaladejo:2016lbb} in green, and the various extractions from the
    amplitudes extracted in Ref.~\cite{Moir:2016srx} in yellow. The pole locations from 
    Fit~1\_4L, Fit~2\_4L, Fit~3\_4L and Fit~4\_4L are shown in red, blue, black and light blue 
    respectively. The green and blue vertical dashed lines represent the $D\eta$ and
    $D_s\Bar{K}$ thresholds, respectively.
    }
    \label{fig:plocRS221all}
\end{figure}

From every volume we use the lowest four energy levels (thus the addition \_4L to
the fit names) of the [000]~$A^+_1$ irreducible representation,\footnote{We did not implement the discretization of our amplitude for moving frame data, since this is technically a lot more demanding and the usefulness of the SU(3) symmetry constraint can already be demonstrated with the rest frame fits.}
where the $S$-wave component gives the dominant contribution.
In the next subsection we discuss the fit results
for Fit~4\_All that was performed 
including all the rest frame lattice
levels.
The obtained parameters for the different fits as well as the $\chi^2$ values found
are listed in Table~\ref{tab:fitparas}. 
Figure~\ref{fig:elvlscomb} shows the energy levels obtained from the fits together 
with the data points in the lattice rest frame.
The uncertainties of
the fit parameters quoted are from the MIGRAD routine of MINUIT.  
Further, using the parameters from the fit, the poles
of the $T$-matrix  in the continuum on the different
Riemann sheets are extracted. 
The resulting pole positions can
be found in Table~\ref{tab:polessu3}. 
It turns out that, contrary to the pole extraction employing
Eq.~(\ref{eq:Kdef}), now in all fits the higher mass pole 
has a mass of about 2.5~GeV and is thus located close
the $D\eta$ and $D_s\bar K$ thresholds (see Fig.~\ref{fig:plocRS221all})

A more detailed comparison of the
performance of the fits shows that for $[\bar 3]$ both
a pole term and a constant term are needed  to obtain an acceptable fit. Moreover, the uncertainties for
the pole parameters that emerged from Fit~1\_4L are
a lot larger than for the other fits (in addition the fit even allows for an additional level very close to the fitting
range).
We therefore exclude both Fit~1\_4L and Fit~2\_4L from  further discussions.
On the other hand, fits of comparable quality emerge, if either
the constant term in the $[6]$ (Fit~3\_4L) or the pole term in the $[6]$ (Fit~4\_4L)
is abandoned. In the latter case the higher pole is generated via the unitarization. 
Note that in our case the number of parameters connected
to the bound state pole is in any case 2 (one coupling constant and a mass), while in case of the fits performed by the  Hadron Spectrum Collaboration this number is 4, for there
an individual coupling is needed for each channel.
The difference in the number
of parameters needed for the $[\bar 3]$ and the $[6]$
channels can be understood straightforwardly from the observation that
the pole in the $[\bar 3]$ is a bound state and further parameters are needed to
obtain a decent fit of the additional energy levels. The pole originated in the $[6]$, 
on the other hand, sits rather high up in the spectrum, having a larger imaginary part, and thus naturally controls
all energy levels that have a sizable contribution from this representation;
either the bare pole or the constant term in $[6]$ provides a seed for the $[6]$ pole.

{
Table~\ref{tab:polessu3} also shows that in all fits  poles appear on RS222 above
the $D_s \bar K$ threshold, however, with large uncertainties
especially on the mass parameters. Since in this energy range RS222 connects directly
to the physical sheet, these poles show up as peaks in the amplitudes at high energies.
Note that analogous poles were also present in the fits performed in the course of the
analysis of Ref.~\cite{Moir:2016srx}
 and in the UChPT amplitude (see Table~\ref{tab:poles}), however, 
they appeared at significantly higher energies.
In Fit~1\_4L this pole can appear quite close to the energy region of interest, given the
large uncertainty in the mass parameter. We interpret this phenomenon as reflecting 
a too large number of parameters in the fit. In the two best fits, namely Fit~3\_4L and Fit~4\_4L, on the other
hand, the poles on RS222 are typically located deep inside the complex plane or rather high above
the threshold, respectively, although within
uncertainties it can appear rather close to
threshold also for Fit~3\_4L, with leads
to the strong rise very close to the higher thresholds.
The large spread in the amplitudes above the $D_s \bar K$ threshold visible in Figs.~\ref{fig:ampcomb}
reflects the bad determination of the highest pole from
the lattice data included in the fits.
}

\begin{figure}
    \centering
    \includegraphics[scale=0.44]{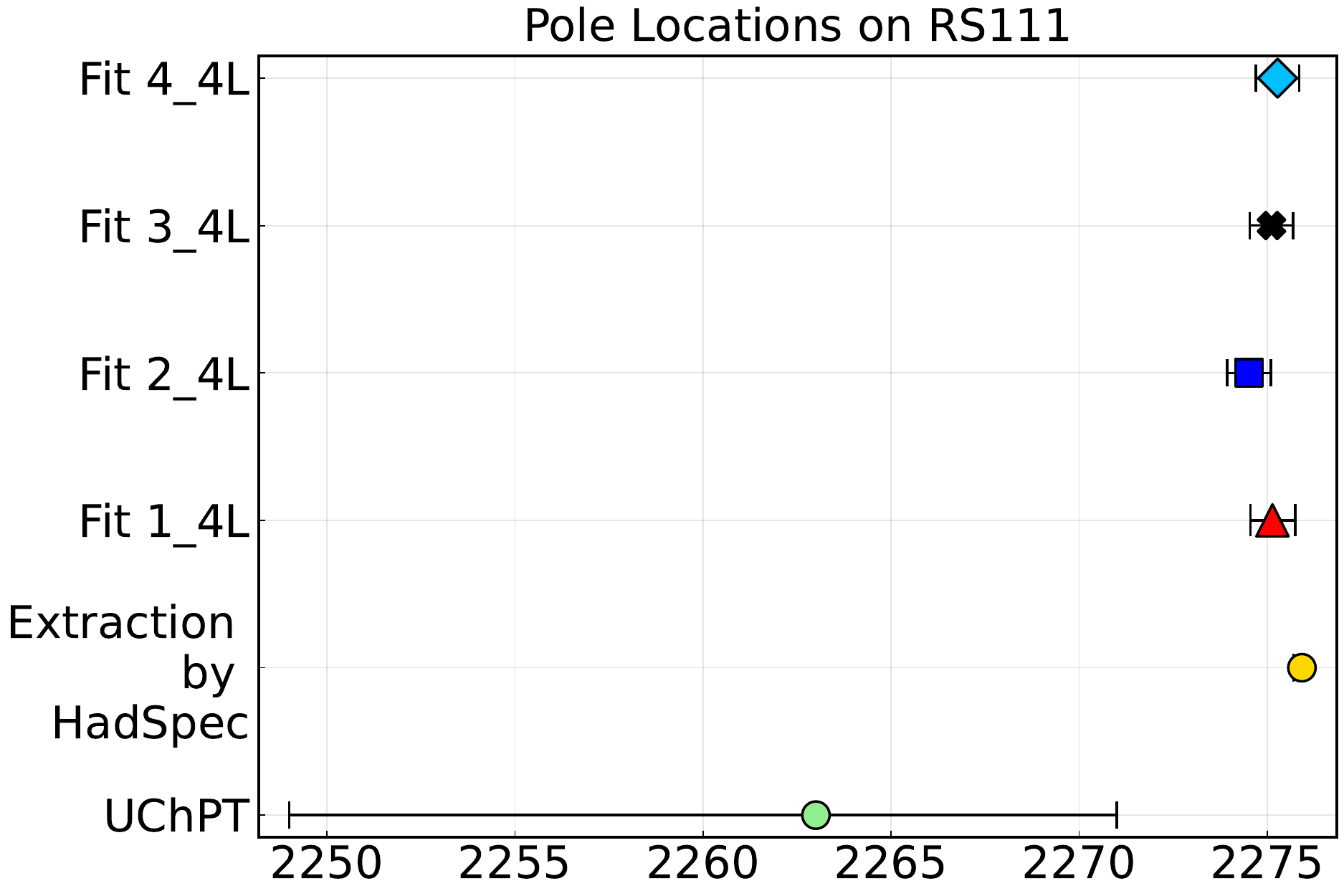}
    \caption{Locations of the RS111 poles from the different fits together with the RS111 pole in
    the UChPT amplitude~\cite{Albaladejo:2016lbb} and from HadSpec~\cite{Moir:2016srx}. 
    }
    \label{fig:plocRS111}
\end{figure}

\begin{figure*}[h]
\centering
    \includegraphics[width=0.45\linewidth]{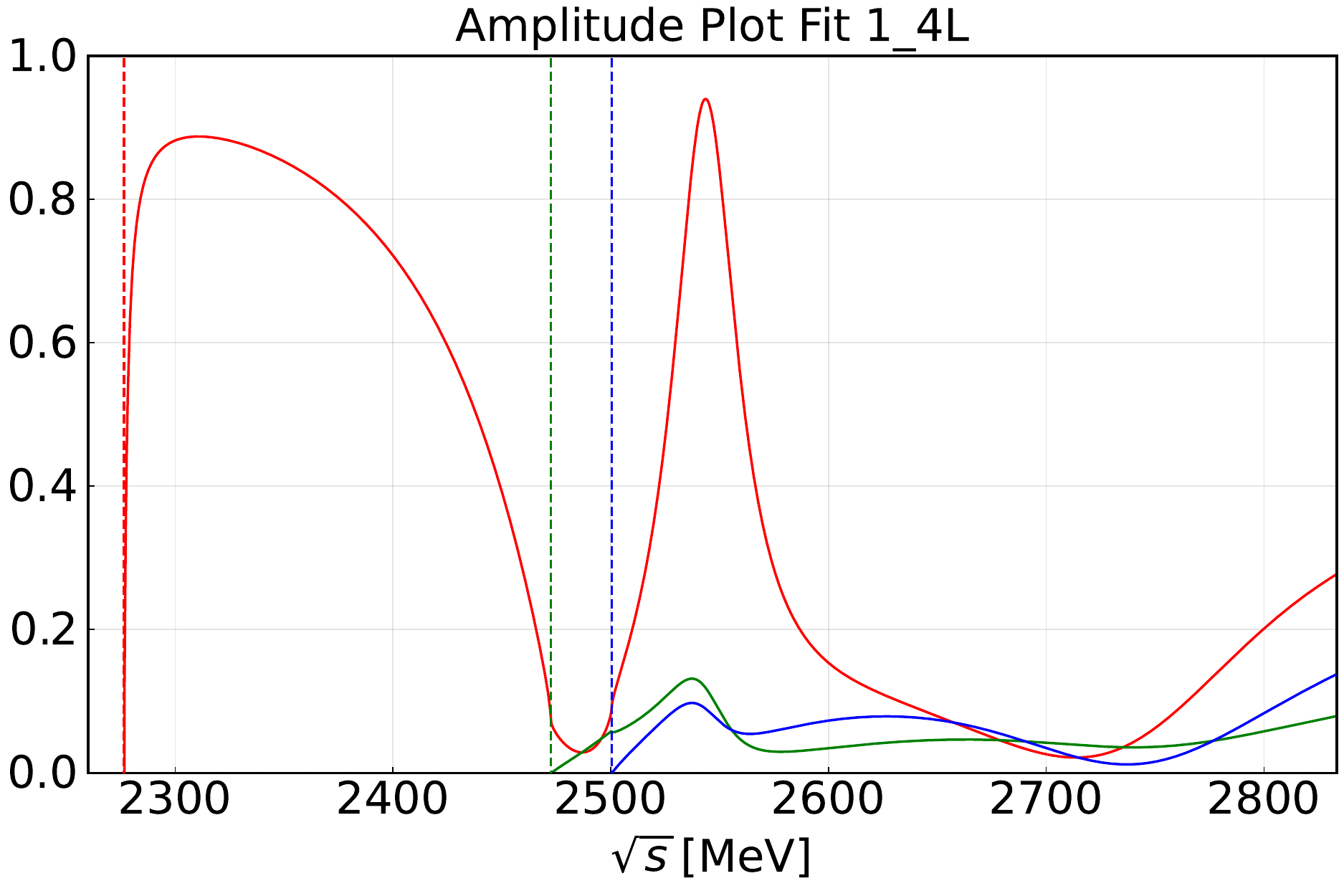}
    \includegraphics[width=0.45\linewidth]{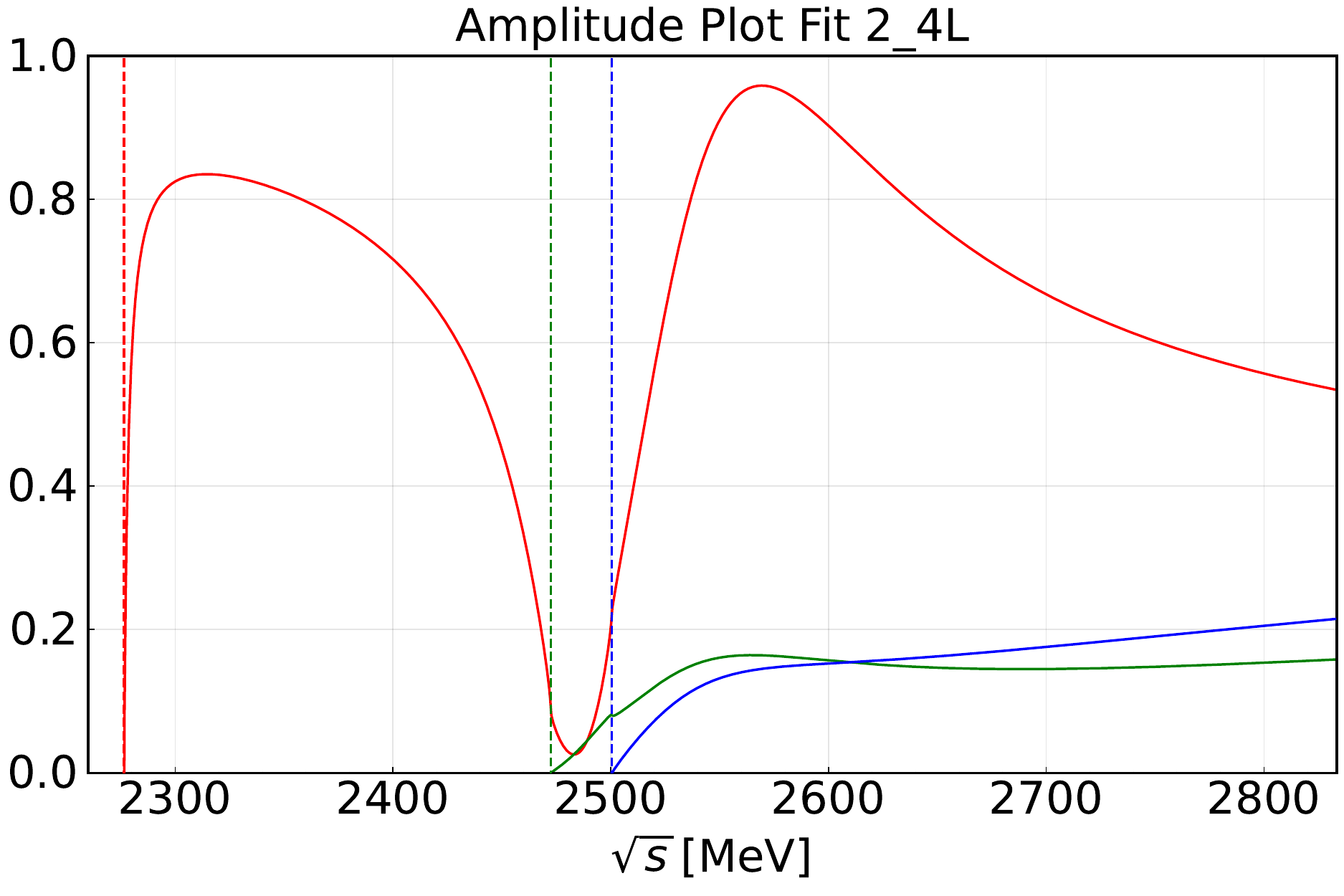}
    \includegraphics[width=0.45\linewidth]{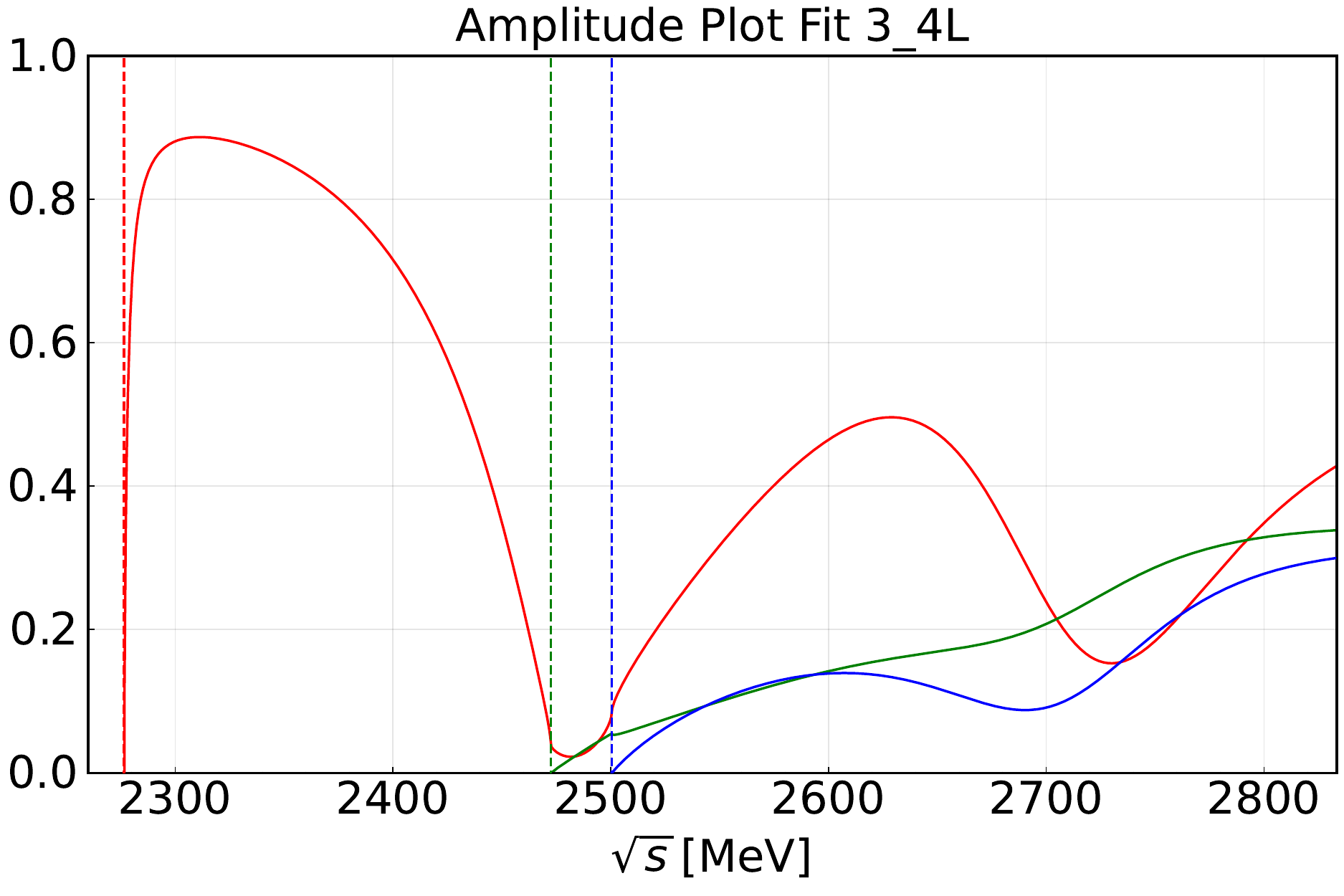}
    \includegraphics[width=0.45\linewidth]{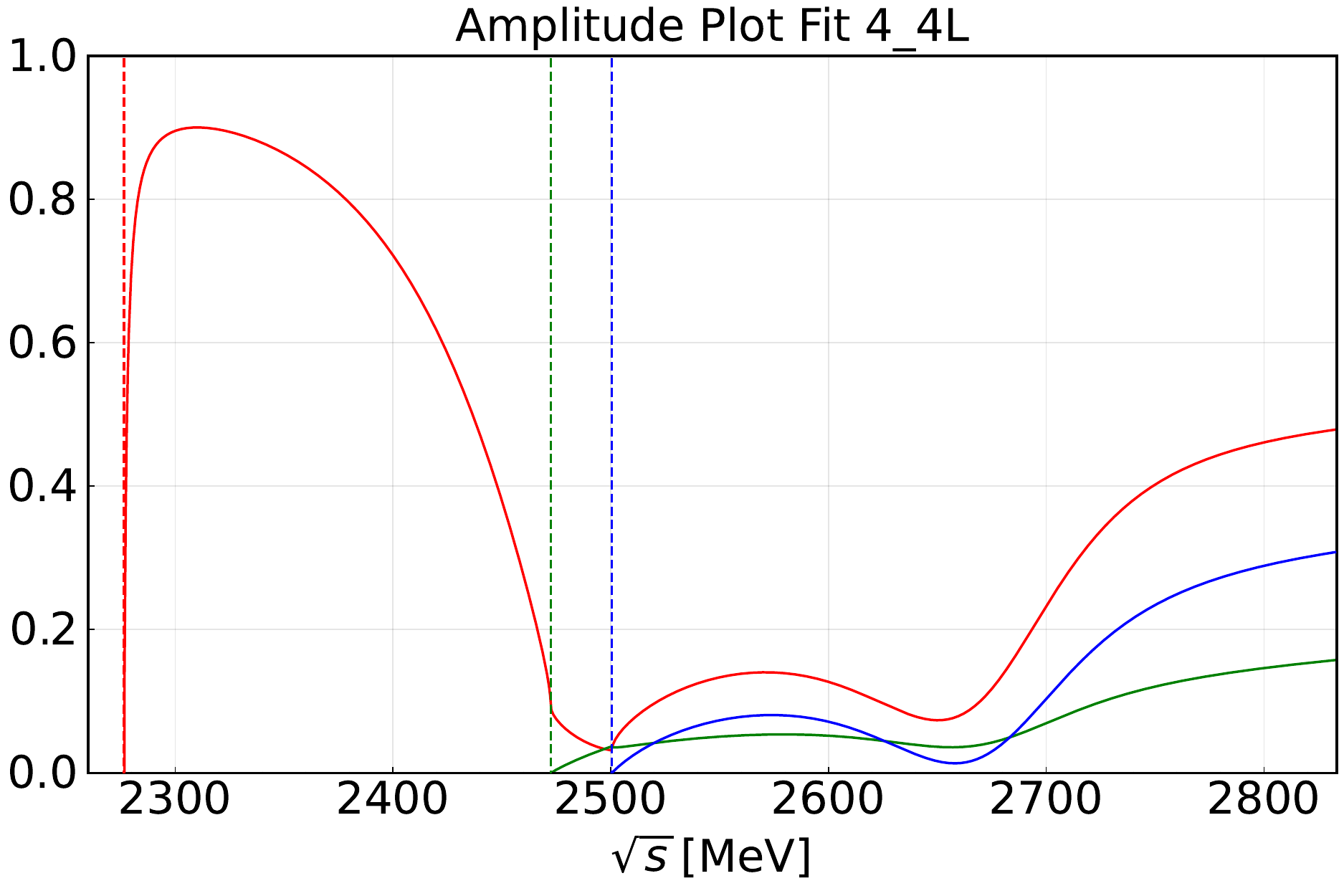}
\caption{The resulting amplitudes from the various fits in the form of $\rho^2 |T|^2$. 
    The $D\pi\text-D\pi, D\eta\text-D\eta$\ and $D_s\Bar{K}\text-D_s\Bar{K}$\ amplitudes are 
    shown in red, green and blue, respectively. The vertical red, green and blue dashed lines show the $D\pi$, $D\eta$\ and the $D_s\Bar{K}$\ thresholds, respectively. The 
    error bands cover the 1$\sigma$\ statistical uncertainties.
   }
    \label{fig:ampcomb}
\end{figure*}

A comparison of the RS111 pole locations from the SU(3) fits
just reported and that the UChPT amplitude~\cite{Albaladejo:2016lbb} and Ref.~\cite{Moir:2016srx} is 
shown in Fig.~\ref{fig:plocRS111}. As expected the location of this bound state pole
is consistent amongst all extractions. 

Note that in all fits the constant term in the $[\overline{15}]$ representation turns out to be repulsive,
in line with the expectations from leading order chiral perturbation theory. This is a
nice and in fact non-trivial confirmation of the hypothesis that, even at pion masses as
high as 391~\mev, already leading order chiral perturbation theory provides valuable guidance for
the physics that leads to the emergence or not appearance of hadronic molecules.

We also tested if we can fit the lattice data when replacing the pole term in the
$[6]$ representation by a pole in the $[\overline{15}]$. Those fits, however, did not converge
and are therefore not reported in the figures and tables. 
The amplitudes arrived from the fits are shown in Fig.~\ref{fig:ampcomb}, there, however,
at significantly higher energies. The appearance of these poles in a direct consequence of
the $K$-matrix parametrization employed.
Moreover, in Fit~1\_4L and Fit~2\_4L those poles are rather close to the threshold 
\begin{table*}
    \centering
    \caption{The best fit values arrived in the 
different fits to the lattice energy levels of Ref.~\cite{Moir:2016srx} based on
Eq.~(\ref{eqn:su3k}), 
along with their $\chi^2/\mbox{dof}$. The symbol '-' is used for parameters
set to zero (or absent) in the particular fit.} 
    \resizebox{\textwidth}{!}{%
    \begin{tabular*}{\textwidth}{@{\extracolsep{\fill}}*{11}{c}@{}}
    \hline\noalign{\smallskip}
        & $g_3$ [GeV] & $m_{\Bar{3}}$ [MeV] & $g_6$ [GeV] & 
        $m_6$ [MeV] & $c_{\Bar{3}}$ & 
        $c_6$ & $c_{\overline{15}}$ & $\chi^2$ & $\chi^2/\mbox{dof}$\\      
        \noalign{\smallskip}\hline\noalign{\smallskip}
         Fit~1\_4L  & $2.92\pm 0.39$ & $2275.1\pm 0.6$ & $0.32\pm 0.32$ & $2542\pm 50$ & $4\pm 3$ &  $0.7\pm 0.4$ &  $-0.6\pm 0.2$ & 7.1 & $1.4$\\
         Fit~2\_4L & $2.31\pm 0.14$ & $2274.5\pm 0.8$ & $0.66\pm 0.17$ & $2560\pm 37$ & - & - & $-0.6\pm 0.2$ & 14 & $1.9$\\
         Fit~3\_4L & $2.91\pm 0.39$ & $2275.1\pm 0.6$ & $1.20\pm 0.62$ & $2735\pm 266$ & $4\pm  2$ & - & $-0.6\pm 0.2$ & 7.3 & $1.2$\\
        Fit~4\_4L & $3.16\pm  0.38$ & $2275.3\pm  0.6$ & - & - & $5\pm  2$ & $1.0\pm  0.2$ & $-0.4\pm  0.2$ &  8.2 &$1.2$\\
        \hline
        Fit~4\_All & $2.4\pm  0.2$ & $2274.8\pm  0.6$ & - & - & $1.1\pm  0.4$ & $0.54\pm  0.06$ & $-0.26\pm  0.09$ &  $29.6$ & $2.1$\\
         \noalign{\smallskip}\hline
    \end{tabular*}}
    \label{tab:fitparas}
\end{table*}

\begin{table*}
    \centering
    \caption{The pole locations from the different fits.}
\begin{tabular*}{\textwidth}{@{\extracolsep{\fill}}ccccc@{}} 
            \hline\noalign{\smallskip}
        Fits & RS111 & RS211 & RS221 & RS222  \\ 
            \noalign{\smallskip}\hline\noalign{\smallskip}
       Fit~1\_4L & $2275.1^{+0.6}_{-0.6} -0\mathit{i}$ & $2515^{+88}_{-19}-23^{+19}_{-88}\mathit{i}$ & $2476^{+136}_{-109}-253^{+181}_{-120}\mathit{i}$  & $2544^{+151}_{-46}-18^{+18}_{-69}\mathit{i}$ \   \\ 
     Fit~2\_4L & $2274.5^{+0.8}_{-0.7} -0\mathit{i}$ & $2498^{+9}_{-10}-20^{+7}_{-6}\mathit{i}$  & $2503^{+12}_{-13}-42^{+19}_{-22}\mathit{i}$ & $2518^{+19}_{-21}-63^{+31}_{-44}\mathit{i}$ \ \\ 
     Fit~3\_3L & $2275.1^{+0.6}_{-0.6} -0\mathit{i}$ & $2512^{+22}_{-67}-50^{+37}_{-20}\mathit{i}$ & $2479^{+41}_{-50}-128^{+103}_{-38}\mathit{i}$ & $2571^{+250}_{-135}-314^{+265}_{-84}\mathit{i}$ \ \\ 
      Fit~4\_4L & $2275.3^{+0.6}_{-0.6} -0\mathit{i}$ & $2518^{+28}_{-17}-92^{+18}_{-28}\mathit{i}$ & $2407^{+59}_{-40}-241^{+43}_{-50}\mathit{i}$ & $2673^{+94}_{-44}-61^{+19}_{-47}\mathit{i}$ \  \\ 
 \hline 
      Fit~4\_All & $2274.8^{+0.6}_{-0.6} -0\mathit{i}$ & 
      $2681^{+46}_{-33}-263^{+43}_{-51}\mathit{i}$ & $2516^{+71}_{-60}-479^{+38}_{-50}\mathit{i}$ & $3123^{+144}_{-99}-359^{+86}_{-162}\mathit{i}$ \  \\ 
            \noalign{\smallskip}\hline
        \end{tabular*}
    \label{tab:polessu3}
\end{table*}

\begin{figure*}
    \centering
    \includegraphics[width=0.33\textwidth]{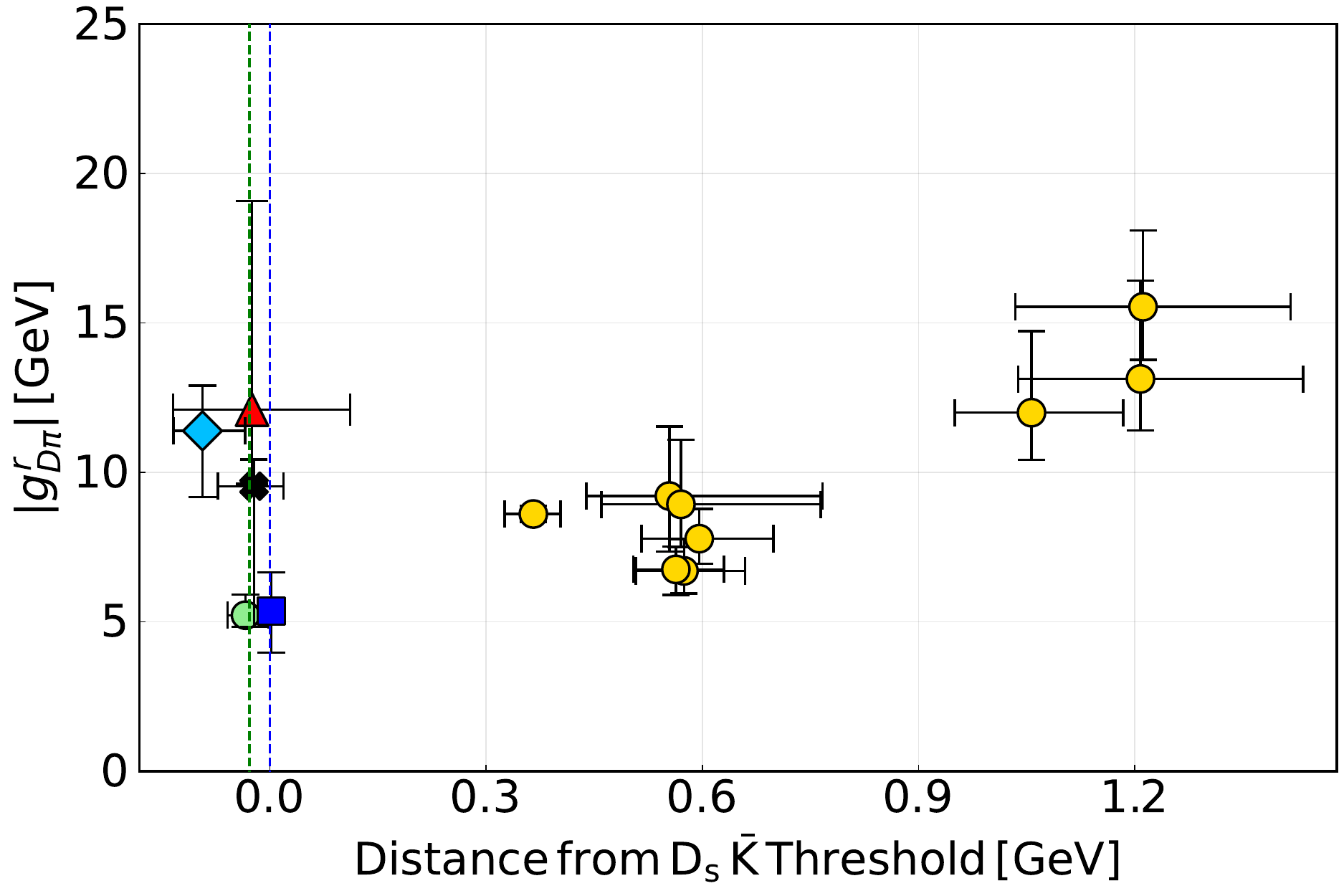}
    \includegraphics[width=0.33\textwidth]{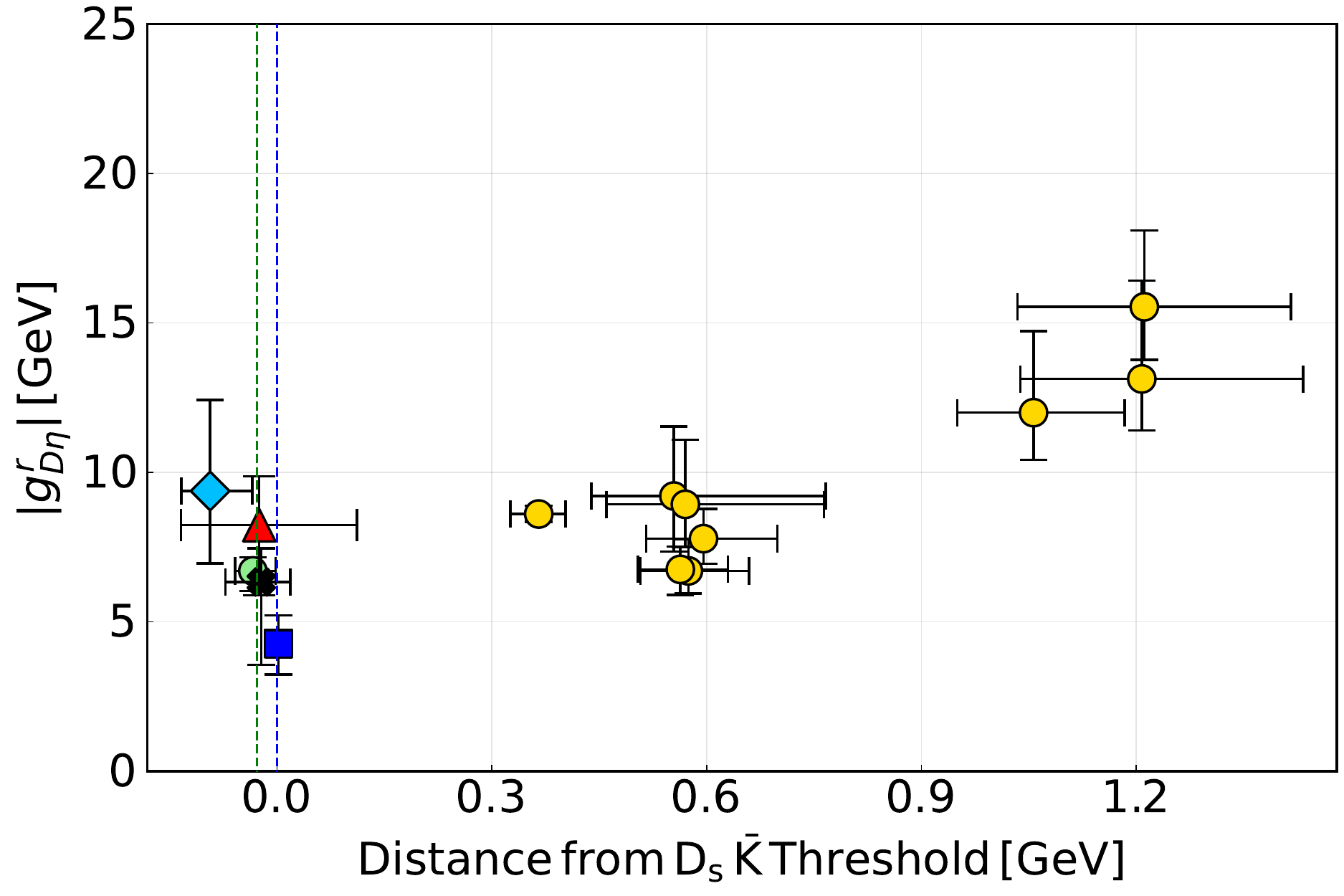}
    \includegraphics[width=0.33\textwidth]{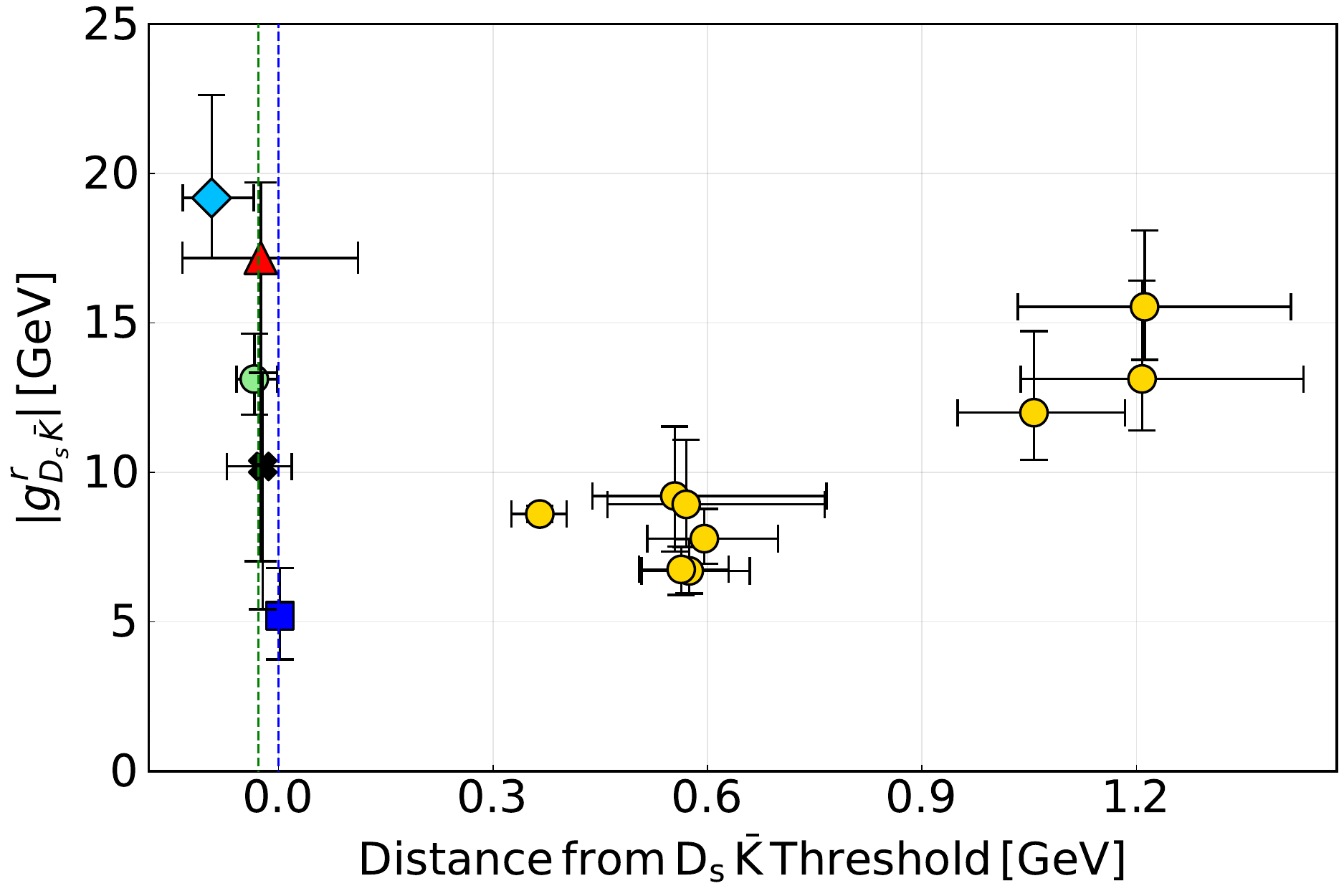}
    \caption{The distance of the real part of the pole on RS221 from the $D_s\Bar{K}$\ 
    threshold versus the effective coupling of the pole to the $D\pi$\ channel (left),  
    $D\eta$\ channel (centre) and  $D_s\bar K$ channel (right). The data points from 
    Fit~1\_4L, Fit~2\_4L, Fit~3\_4L and Fit~4\_4L are shown in red, blue, black and light blue, 
    respectively. Those for the amplitudes obtained in Ref.~\cite{Moir:2016srx} are shown in yellow
    and UChPT amplitude~\cite{Albaladejo:2016lbb} in green. The green 
    and blue vertical dashed lines show the $D\eta$ and $D_s\Bar{K}$\ thresholds, respectively. The error bars 
    represent the 1$\sigma$\ statistical uncertainty.
    }
    \label{fig:SU(3)Res}
\end{figure*}

To complete the discussion, in 
Table~\ref{tab:resu3} we show the square root of the absolute values 
of the RS221 pole residues to the respective 
channels derived from the fits. Further, Fig.~\ref{fig:SU(3)Res} shows the distance of the RS221 pole to the $D_s\bar K$ threshold, as defined in Eq.~\eqref{eqn:pole distance}, versus the effective coupling of the pole to the $D\pi$, $D\eta$ and $D_s\bar K$ channels, respectively. Besides Fit~4\_4L,
all values are statistically consistent with the intercepts arrived at in Section~\ref{subsec:DxR} within errors.

\begin{table}
    \centering
    \caption{The absolute value of the square root of the RS221 pole residues obtained by the 
    SU(3) flavor constrained $K$-matrix to the respective channels.}
    \begin{tabular}
    {@{\extracolsep{\fill}}ccccccc@{}} 
    \hline\noalign{\smallskip}
    Parameter set & $|\gdp|$ [\gev] & $|\gdeta|$ [\gev] & $|\gdsk|$ [\gev] \\
    \noalign{\smallskip}\hline\noalign{\smallskip}
    Fit~1\_4L & $12_{-2}^{+7}$ & $8_{-2}^{+2}$ & $17_{-10}^{+2}$  \\
    Fit~2\_4L & $5_{-1}^{+1}$ & $4_{-1}^{+1}$ & $5_{-2}^{+2}$  \\
    Fit~3\_4L & $10_{-5}^{+1}$ & $6_{-3}^{+1} $ & $10_{-5}^{+3}$  \\
    Fit~4\_4L & $11_{-2}^{+2}$ & $9_{-2}^{+3}$ & $19_{-2}^{+3}$  \\
\hline
    Fit~4\_All & $13.4_{-0.5}^{+0.5}$ & $8.5_{-0.7}^{+0.8}$ & $17.8_{-0.9}^{+0.8}$  \\
  \noalign{\smallskip}\hline
\end{tabular}
    \label{tab:resu3}
\end{table}

\subsection{Inclusion of the Higher Lattice Levels}

Our fit amplitudes are formulated as a momentum expansion.
Accordingly, to find our main results, we performed fits
with including only the lowest 4 lattice levels at each
volume. However, to check for stability of our findings,
we also performed fits with Fit~3 and Fit~4
to all rest frame levels ([000]~$A^+_1$ irrep.) --- the resulting fits are
labeled as Fit~3\_ALL and Fit~4\_All.
It turns out that the former parametrization does not allow
for a decent fit (the best $\chi^2/$dof we can achieve
is 5.5). 
It calls for introducing additional bare poles or/and momentum-dependent contact terms.

\begin{figure*}
    \includegraphics[width=0.5\textwidth]{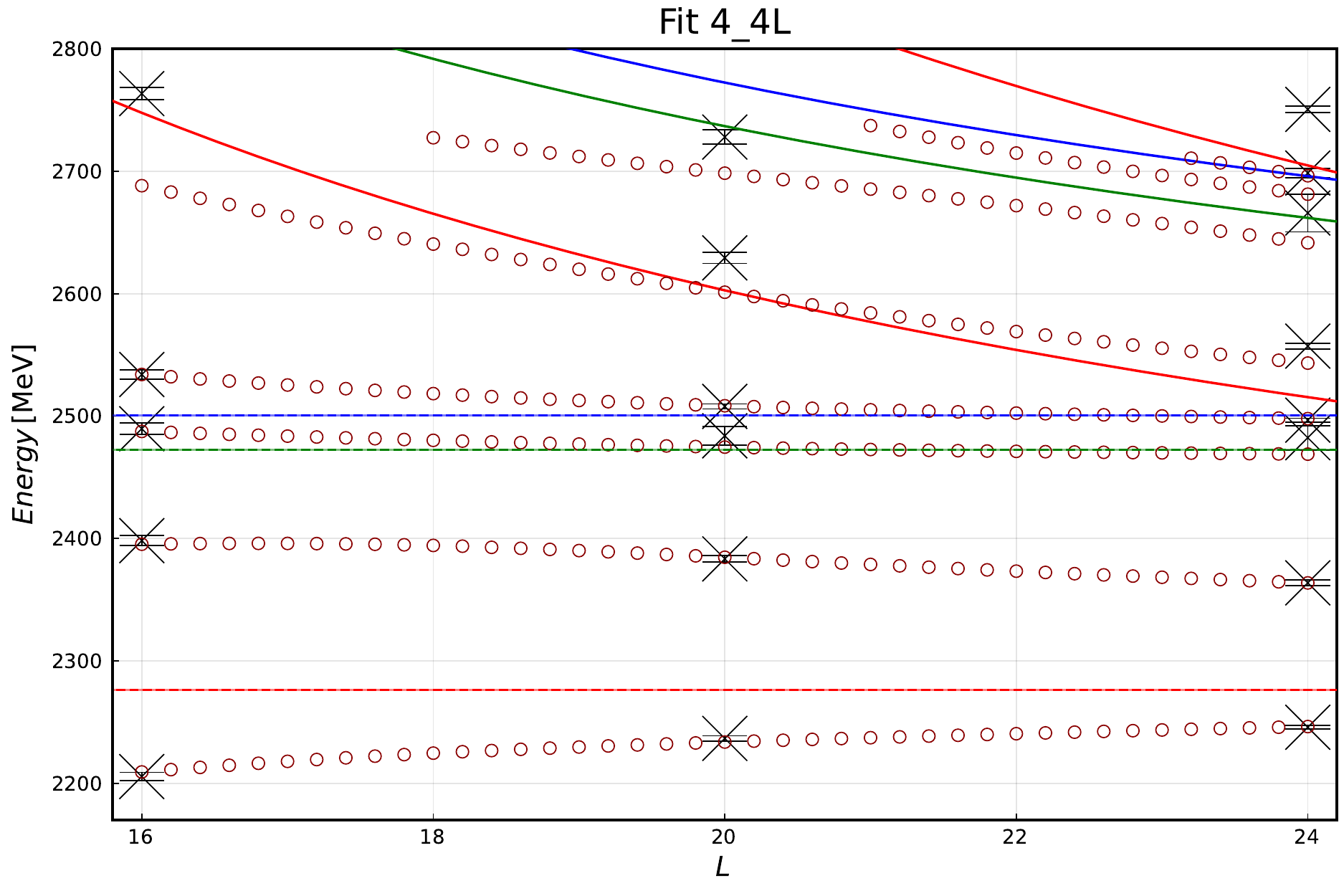}
    \includegraphics[width=0.5\textwidth]{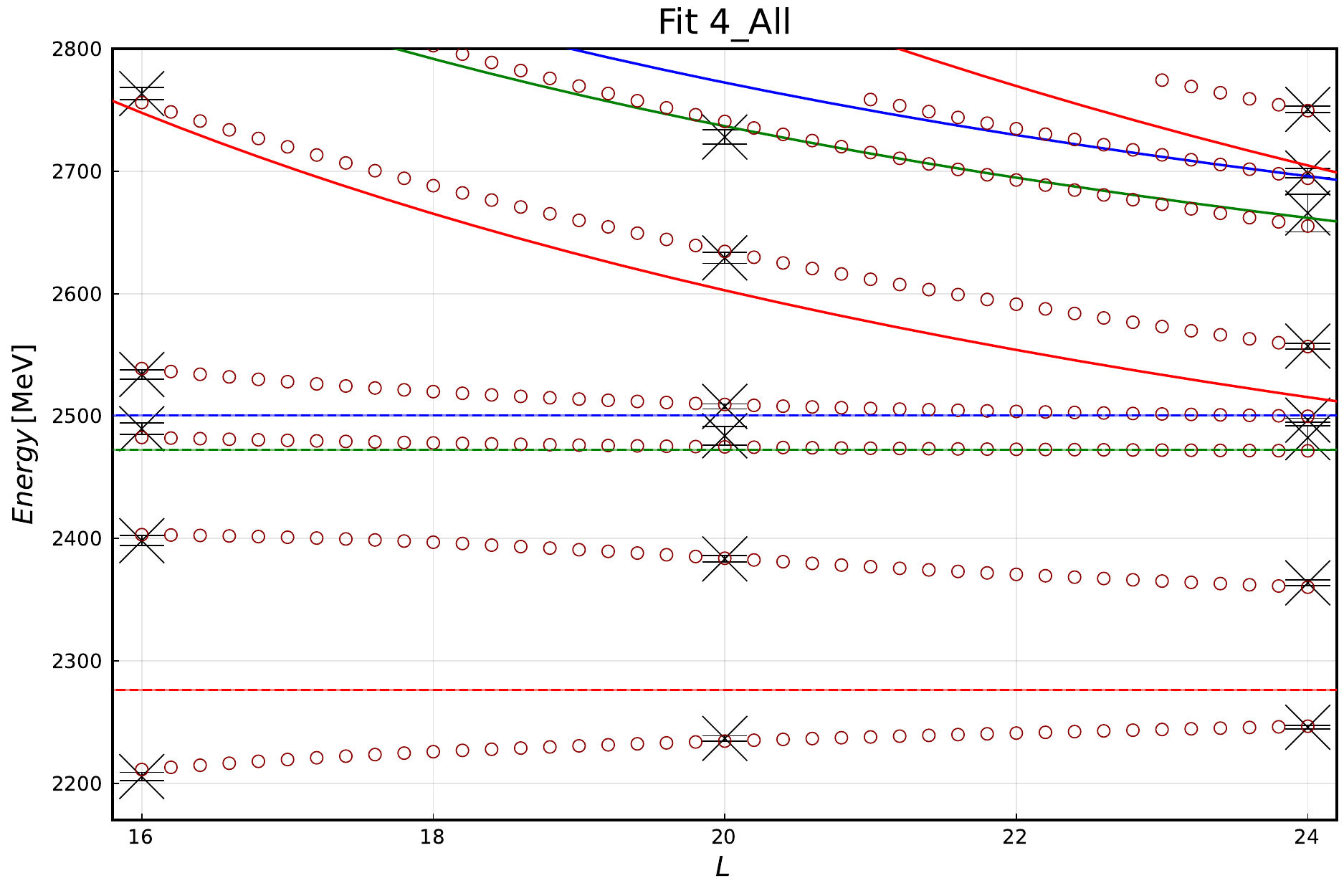}
    \caption{The left panel shows the finite volume energy levels arrived at for Fit~4\_4L (only
    the first four lowest energy levels of each volume
    included in the fit), while the right panel
    shows them for Fit~4\_All. 
    The black crosses are the lattice energy level data. The dark red circles show the energy levels from the amplitudes. The solid lines show the non-interacting energy levels while the dashed line shows the thresholds. The $D\pi$, $D\eta$ and $D_s\Bar{K}$ energies are shown in red, green and blue, respectively.
    }
    \label{fig:energy_curves}
\end{figure*}

\begin{figure}
    \includegraphics[width=0.5\textwidth]{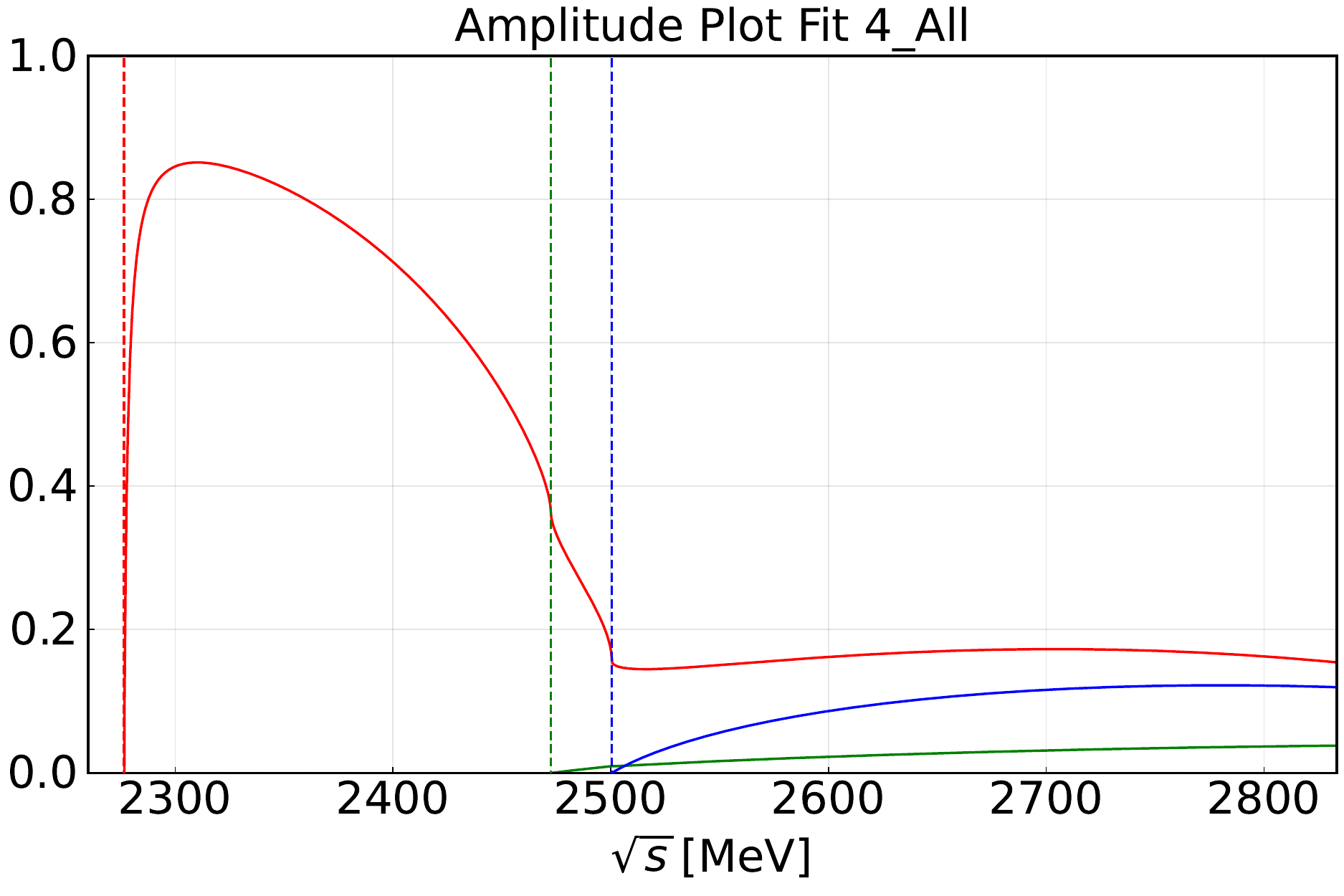}
    \caption{The resulting amplitudes of Fig 4\_ALL in the form of $\rho^2 |T|^2$. 
    The $D\pi\text-D\pi, D\eta\text-D\eta$\ and $D_s\Bar{K}\text-D_s\Bar{K}$\ amplitudes are 
    shown in red, green and blue respectively. The vertical red, green and blue lines show the $D\pi$, $D\eta$\ and the $D_s\Bar{K}$\ thresholds, respectively. The 
    error bands cover the 1$\sigma$\ statistical uncertainties.}
    \label{fig:allfitamp}
\end{figure}

\begin{figure}
    \includegraphics[width=0.45\textwidth]{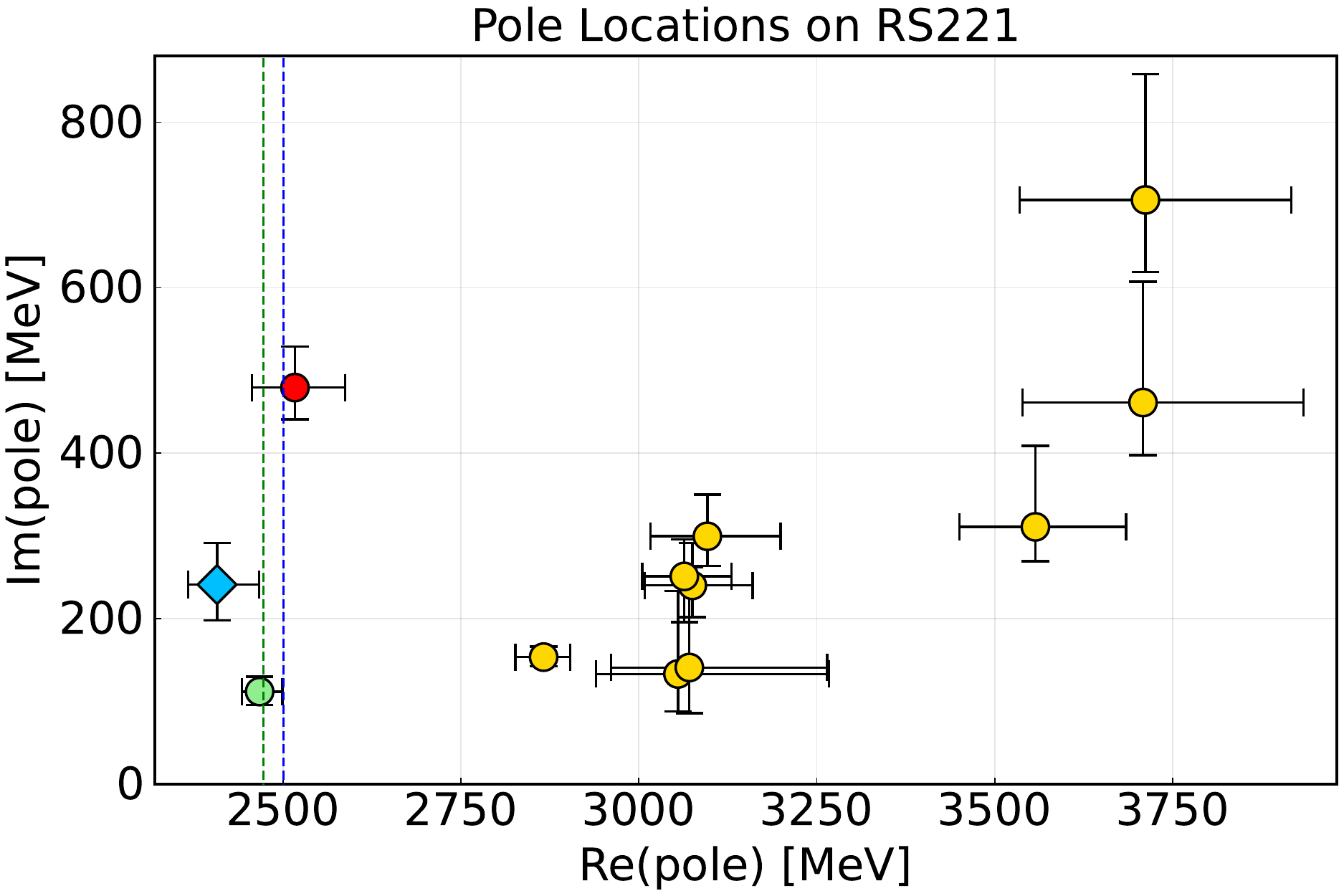}
    \caption{Comparison of locations of the RS221 poles. The pole for Fit~4\_All is shown in red, together with the pole reported
    in UChPT amplitude~\cite{Albaladejo:2016lbb} in green, and the various extractions from the
    amplitudes extracted in Ref.~\cite{Moir:2016srx} in yellow. The pole location from 
    Fit~4\_4L, when including only the lower energy levels, is shown in light blue. 
    The green and blue vertical dashed lines represent the $D\eta$ and $D_s\Bar{K}$ thresholds, respectively.
    }
    \label{fig:poleloccomp}
\end{figure}

In the left panel of Fig.\ref{fig:energy_curves} we show a comparison of the
original fit, Fit~4\_4L, with the full spectrum, in the 
right panel the fit results for Fit~4\_All, where the higher levels are
included in the fit. 
The parameters arrived from the new fit are also  shown in 
Table~\ref{tab:fitparas}. Clearly, the fit is not excellent,
however, when compared to only the rest frame
levels, amplitude 4 and amplitude 6 from Ref.~\cite{Moir:2016srx} reach
 $\chi^2$ values of $36$ and $24$, respectively, 
and are thus of similar quality. 
The amplitude plots arrived from Fit~4\_All are shown in Fig.~\ref{fig:allfitamp}.
Though the bound state pole does not 
change, there is a change in 
the location of the higher pole in sheet RS221, which however has a real part similar to that from Fit~4\_4L. 
The previous pole location (from Fit~4\_4L) and the new pole location found 
from fitting including the higher energy levels are shown in Fig.\ref{fig:poleloccomp}. 
We therefore conclude that it is a stable result from
our analysis that, as soon as SU(3) constraints are included
in the fits, there are always poles close to the
$D_s\bar K$ and $D\eta$ threshold --- we do not find 
anymore the large scatter of the original amplitudes.
   
\section{Summary and Discussion}
\label{sec:summ}

We investigated the pole content of the 
nine $K$-matrix parame\-trizations provided in Ref.~\cite{Moir:2016srx} 
in an analysis of lattice data for open charm states in the $(S=0,I=\frac{1}{2})$ channel. 
In addition to the bound state pole
reported in Ref.~\cite{Moir:2016srx}, in every amplitude additional poles were found 
on unphysical Riemann sheets, however, their locations vary strongly between
the different parametrizations.
On the other hand various investigations employing 
UChPT find that the structure
observed in various experiments in the channel $(S=0,I=\frac{1}{2})$ with
open charm originates from the interplay of two $D_0^*$ poles.
In this paper we explain the origin of this seeming contradiction.
In particular it is shown that also in the lattice
analysis two poles are needed and that, although the
poles scatter so dramatically in location, their effects on the amplitudes were comparable
in all parametrizations. 
This is possible since all poles are located 
on an hidden sheet, such that their effect on the 
scattering amplitude becomes  visible at the threshold. In such a situation
the distance from the threshold can be overcome by an enhanced residue. 
This mechanism was observed before as a general feature of Flatt\'e amplitudes~\cite{Baru:2004xg}.

To discuss the location of the higher pole directly from the lattice data,
we propose to use an amplitude constrained by SU(3) flavor symmetry.
The flavor constrained amplitude well reproduces the energy levels and produces a pole in the RS221 sheet
close to $D\eta$ and $D_s\Bar{K}$ thresholds consistent to that of the UChPT amplitude. 
Such an SU(3) symmetric construction of the $K$ matrix may be used in analyzing other lattice data and also experimental data where multiple channels are involved.

\begin{acknowledgements}
We are very grateful to Christopher Thomas and David Wilson for very valuable discussions and for providing us with the results of Ref.~\cite{Moir:2016srx}.
This work is supported in part by the NSFC and the Deutsche Forschungsgemeinschaft (DFG) through the funds provided to the Sino-German Collaborative Research Center TRR110 ``Symmetries and the Emergence of Structure in QCD'' (NSFC Grant No.~12070131001, DFG Project-ID~196253076); by the Chinese Academy of Sciences under Grant No.~XDB34030000; by the National Natural Science Foundation of China (NSFC) under Grants No.~12125507, No.~11835015, No.~12047503, and No.~11961141012; and by the MKW NRW under the 
funding code NW21-024-A.
The work of UGM was supported further by the Chinese Academy of Sciences
(CAS) President’s International Fellowship Initiative (PIFI) (grant no. 2018DM0034) and the VolkswagenStiftung (grant no. 93562).
\end{acknowledgements}

\appendix
\section{Additional information}
\label{append}

\begin{figure*}
    \centering
    \includegraphics[width=0.49\textwidth]{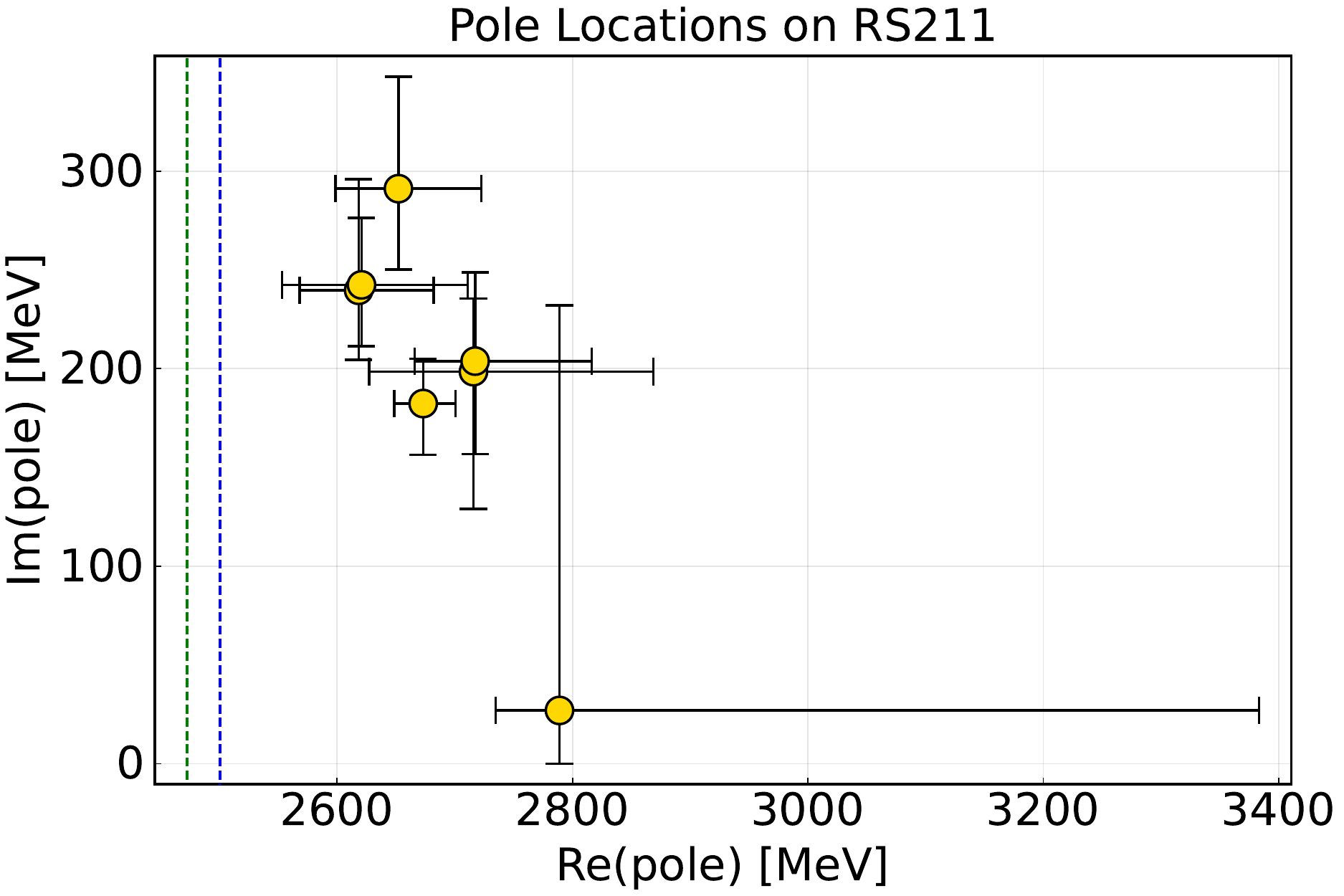} 
    \includegraphics[width=0.49\textwidth]{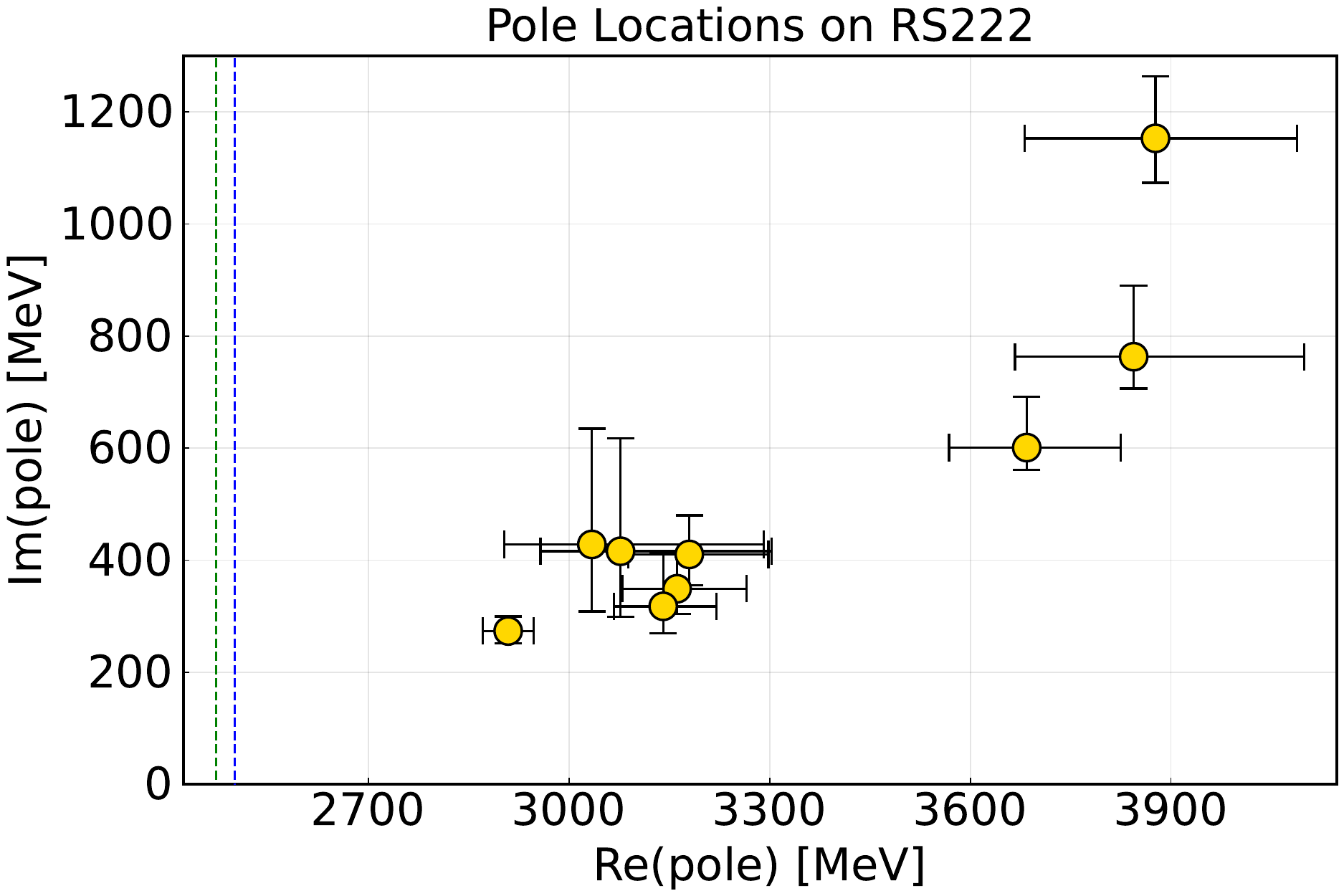}
    \caption{Pole locations on sheet RS211 (left) and on sheet RS222 (right) of the complex energy plane. 
    The vertical green and blue dashed lines represent the $D\eta$ and $D_s\Bar{K}$ thresholds, respectively.
    The error bars show the $1\sigma$\ statistical uncertainty.
    }
    \label{fig:plocRS211RS222}
\end{figure*}

\begin{figure*}
    \centering
    \includegraphics[width=0.33\textwidth]{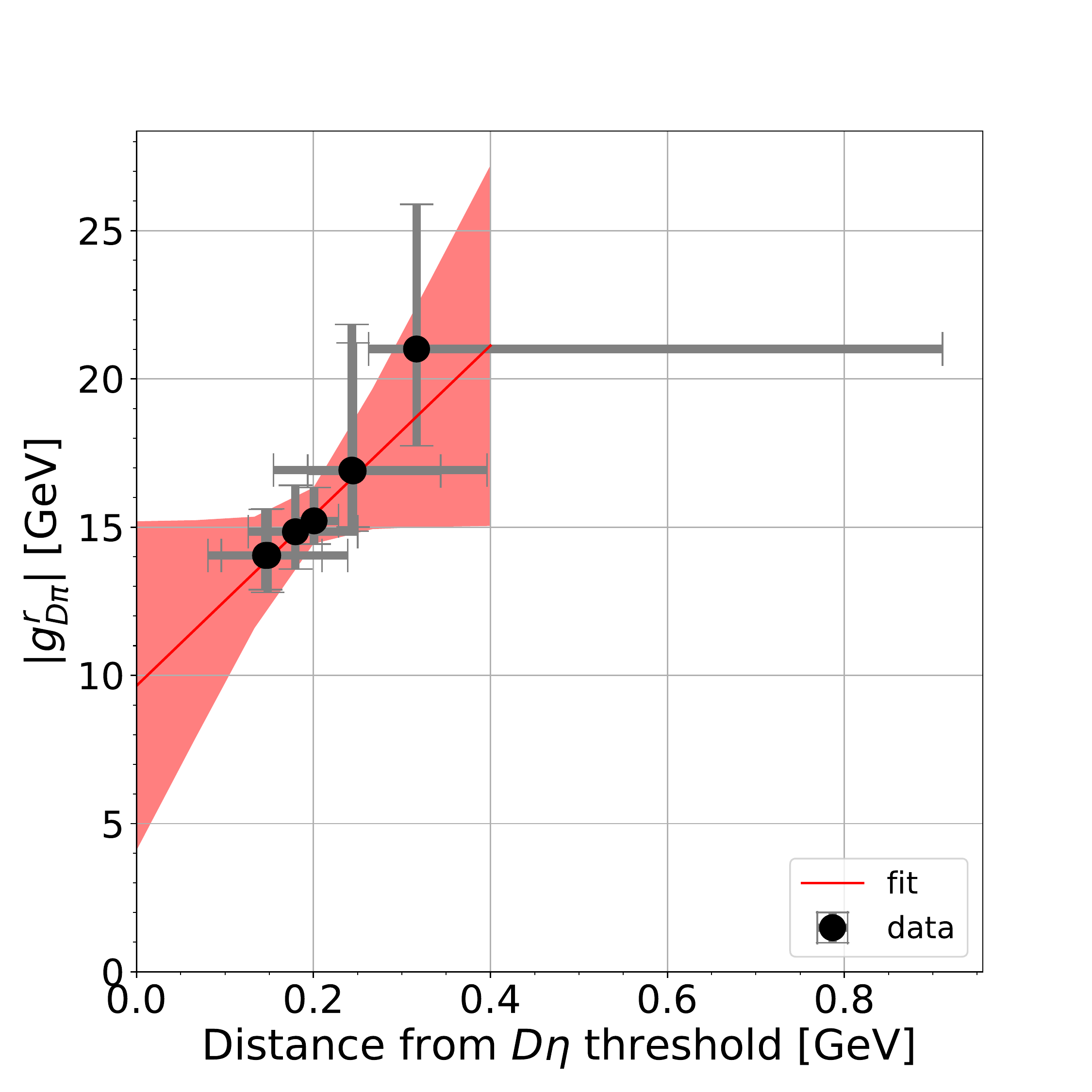}
    \includegraphics[width=0.33\textwidth]{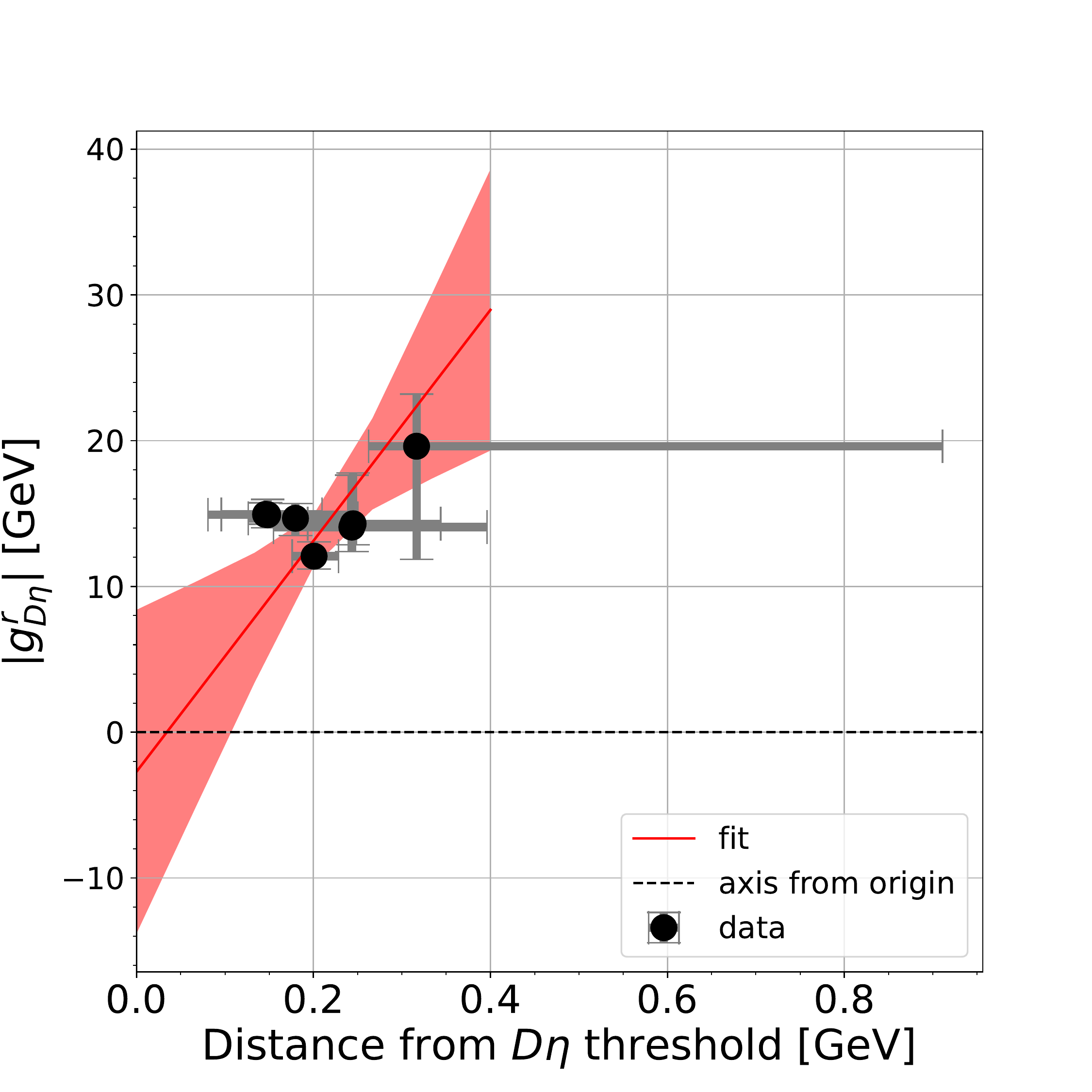}
    \includegraphics[width=0.33\textwidth]{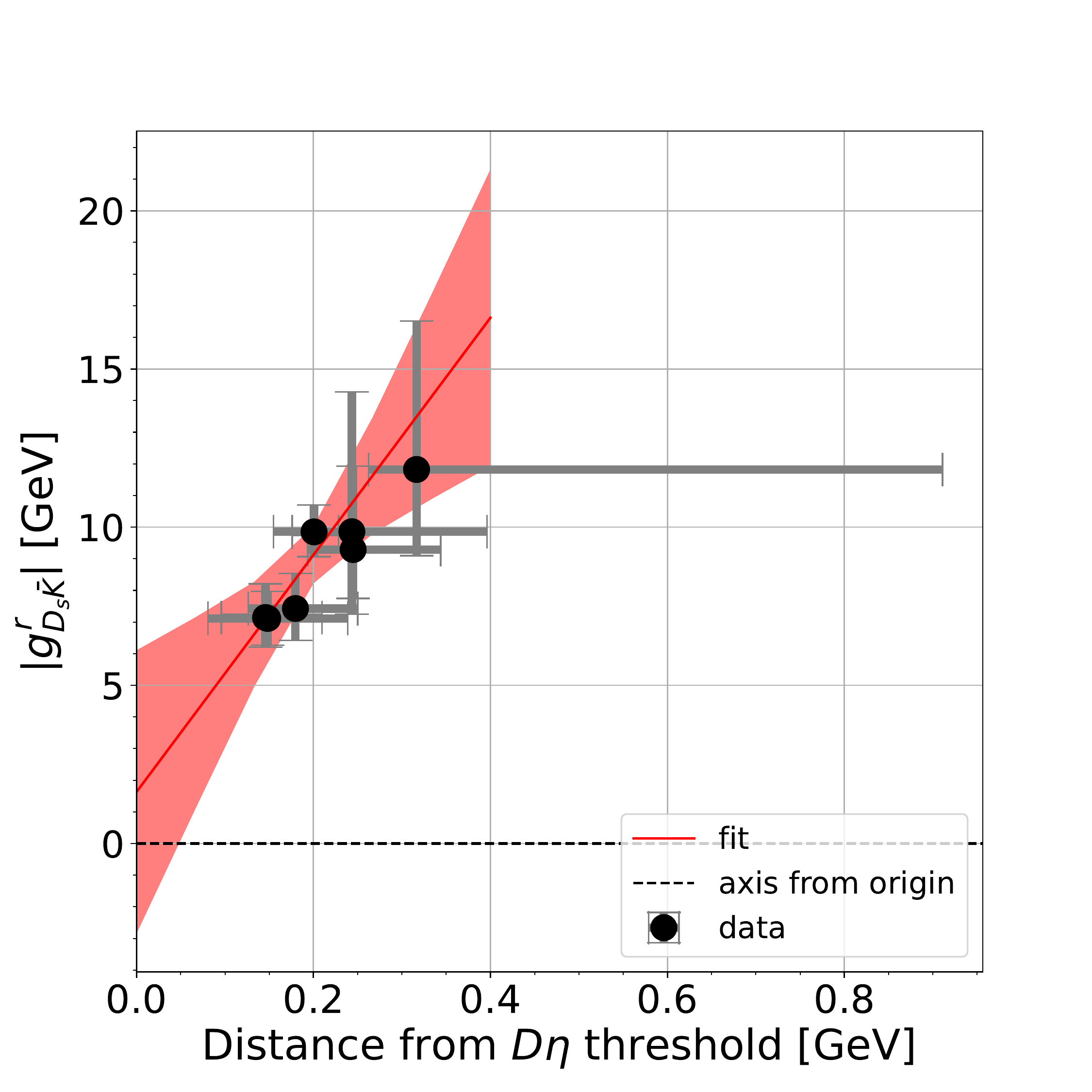}
    \caption{The distance of real part of the pole on sheet RS211 from $D\eta$ 
    threshold versus the effective coupling of the pole to the 
    $D\pi$ channel (left), $D\eta$ channel (center) and $D_s\Bar{K}$ channel (right). The red line shows the 
    straight line fit. The red band encloses the 1$\sigma$ uncertainty of the fit.
    }
    \label{fig:RS211DxC}
\end{figure*}

\begin{figure*}
    \centering
    \includegraphics[width=0.33\textwidth]{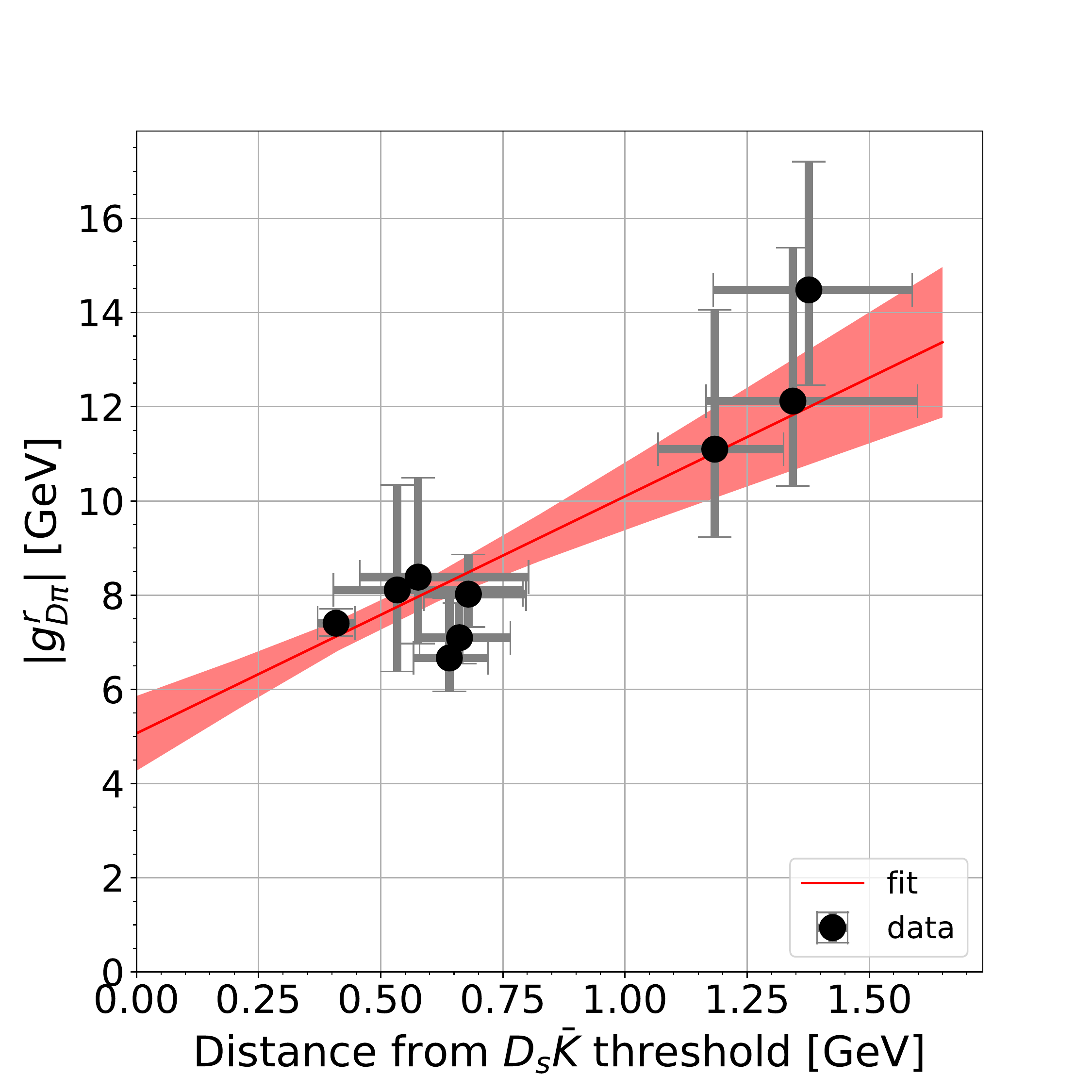}
    \includegraphics[width=0.33\textwidth]{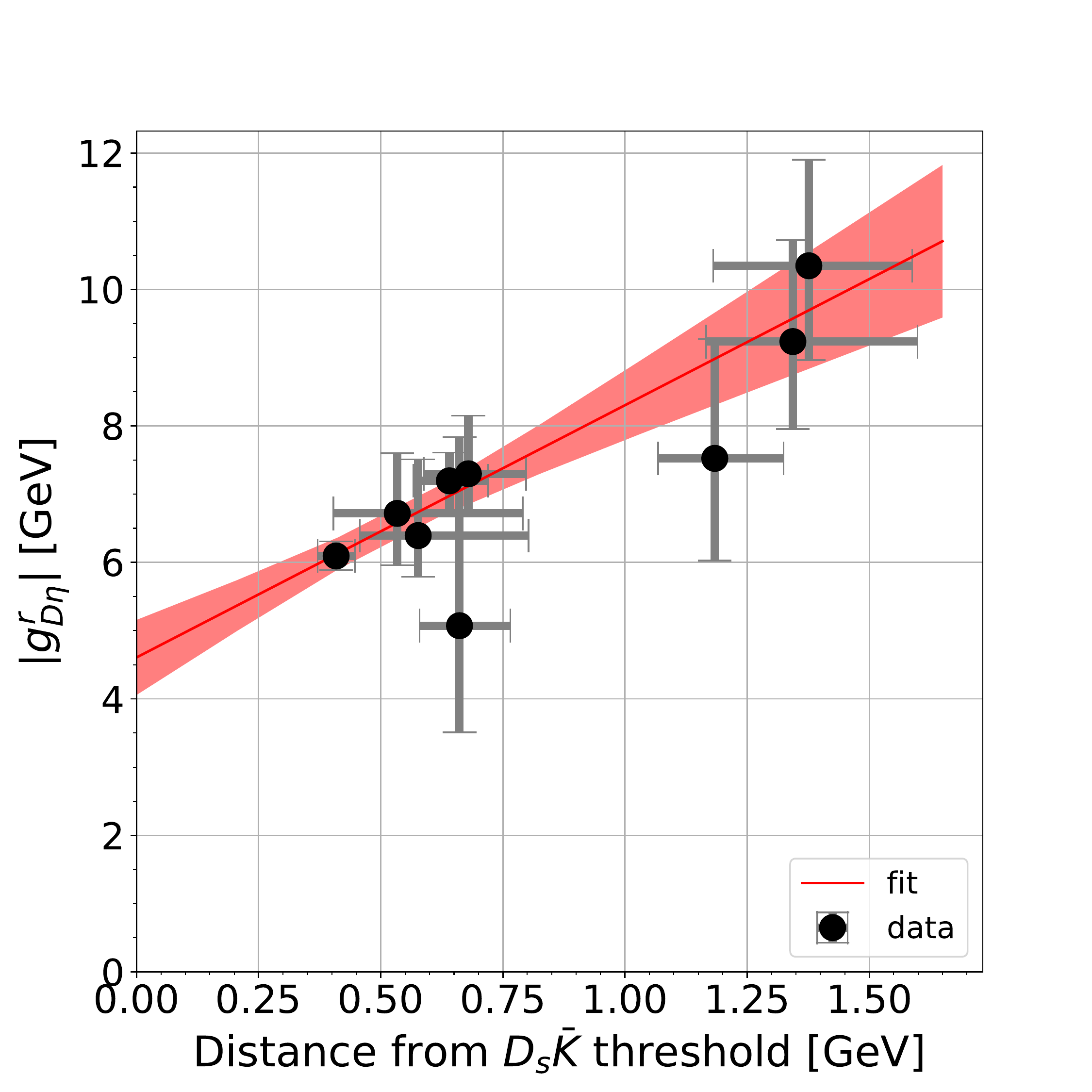}
    \includegraphics[width=0.33\textwidth]{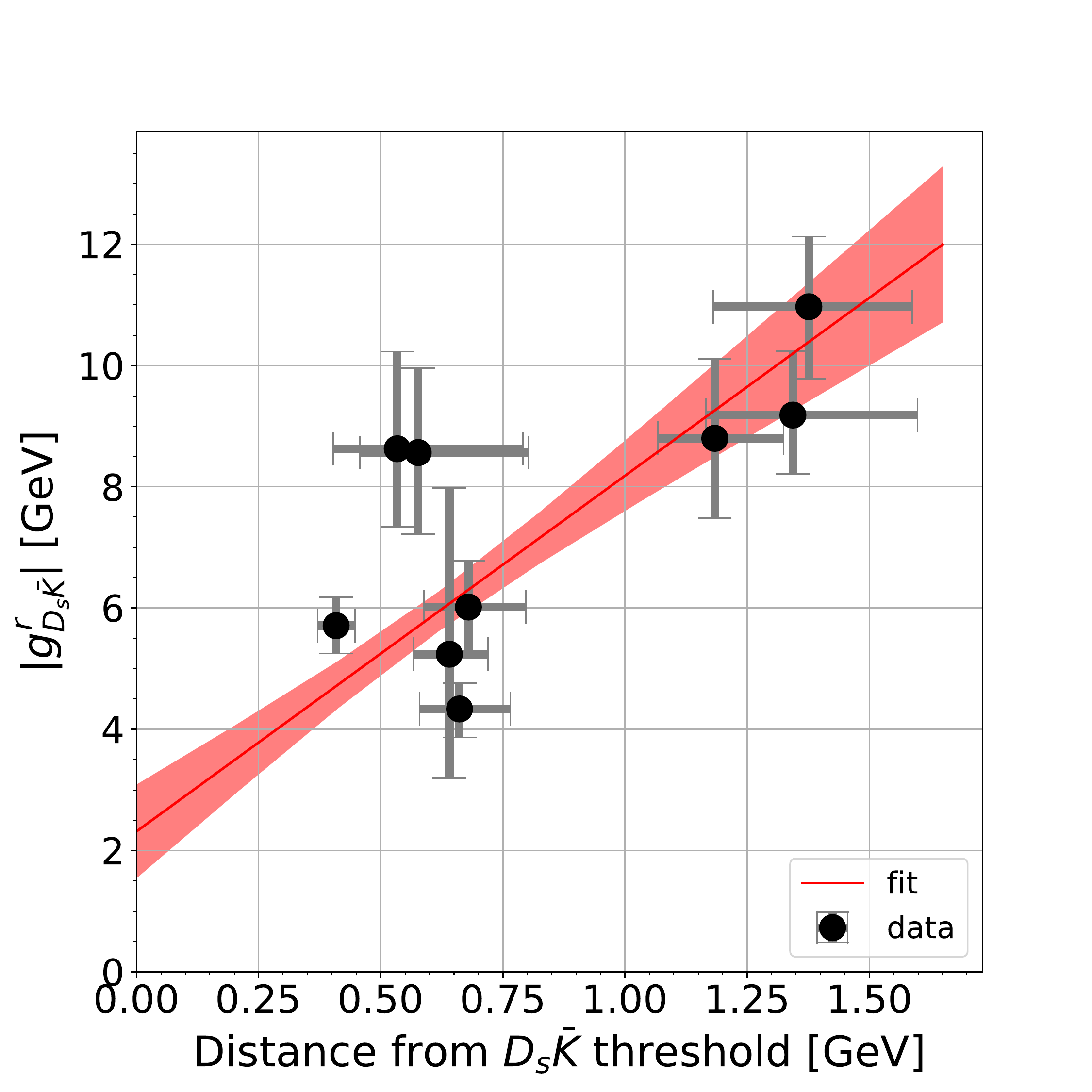}
    \caption{The distance of real part of the pole on sheet RS222 from $D_s\Bar{K}$ 
    threshold versus the effective coupling of the pole to the 
    $D\pi$ channel (left), $D\eta$ channel (center) and $D_s\Bar{K}$ channel (right). The red line shows the 
    straight line fit. The red band encloses the 1$\sigma$ uncertainty of the fit. 
    }
    \label{fig:RS222DxC}
\end{figure*}

In this appendix we provide additional Figs.\ref{fig:plocRS211RS222}-\ref{fig:RS222DxC} 
completing the
presentation. In particular we show the analogous pole analysis for RS221 presented
in the main text for poles on other sheets.  

\bibliography{twopole_paper}

\end{document}